\documentclass[journal,letterpaper,twoside]{IEEEtran}

\ifCLASSINFOpdf
\else
\fi
\usepackage[cmex10]{amsmath}
\usepackage{cite}
\usepackage{mathtools}
\usepackage[pdftex]{graphicx}

\usepackage{color}
\usepackage{soul}
\usepackage{url}

\usepackage[textfont=footnotesize,labelfont=footnotesize]{caption}
\usepackage{subcaption}
\usepackage[caption=false,labelformat=simple]{subfig}

\begin{document}
%
\title{Full-Duplex Transceiver System Calculations: Analysis of ADC and Linearity Challenges}
%
%
%
\author{Dani~Korpi,
        Taneli~Riihonen,~\IEEEmembership{Member,~IEEE,}
        Ville~Syrj\"{a}l\"{a},~\IEEEmembership{Member,~IEEE,}
        Lauri~Anttila,~\IEEEmembership{Member,~IEEE,}
        Mikko~Valkama,~\IEEEmembership{Member,~IEEE,}
        and~Risto~Wichman\vspace{-7mm}
\thanks{Manuscript received June 5, 2013; revised December 18, 2014; accepted March 22, 2014. The published version of this article is available at http://dx.doi.org/10.1109/TWC.2014.2315213}%
\thanks{D. Korpi, V. Syrj\"{a}l\"{a}, L. Anttila, and M. Valkama are with the Department
of Electronics and Communications Engineering, Tampere University of Technology, PO Box 692, FI-33101, Tampere, Finland, e-mail: dani.korpi@tut.fi.}
\thanks{T. Riihonen  and R. Wichman are with the Department of Signal Processing and Acoustics, Aalto University School of Electrical Engineering, PO Box 13000, FI-00076, Aalto, Finland.}%
\thanks{The research work leading to these results was funded by the Academy of Finland (under the projects \#259915, \#258364 "In-band Full-Duplex MIMO Transmission: A Breakthrough to High-Speed Low-Latency Mobile Networks"), the Finnish Funding Agency for Technology and Innovation (Tekes, under the project "Full-Duplex Cognitive Radio"), the Linz Center of Mechatronics (LCM) in the framework of the Austrian COMET-K2 programme, and Emil Aaltonen Foundation.}%
\thanks{\textcopyright \hspace{1pt} 2014 IEEE. Personal use of this material is permitted. Permission from IEEE must be obtained for all other uses, in any current or future media, including reprinting/republishing this material for advertising or promotional purposes, creating new collective works, for resale or redistribution to servers or lists, or reuse of any copyrighted component of this work in other works.}
}

\markboth{IEEE TRANSACTIONS ON WIRELESS COMMUNICATIONS}%
{Korpi \lowercase{\textit{et al.}}: Full-Duplex Transceiver System Calculations: Analysis of ADC and Linearity Challenges}
%


\maketitle

\begin{abstract}

Despite the intensive recent research on wireless single-channel full-duplex communications, relatively little is known about the transceiver chain nonidealities of full-duplex devices. In this paper, the effect of nonlinear distortion occurring in the transmitter power amplifier (PA) and the receiver chain is analyzed, alongside with the dynamic range requirements of analog-to-digital converters (ADCs). This is done with detailed system calculations, which combine the properties of the individual electronics components to jointly model the complete transceiver chain, including self-interference cancellation. They also quantify the decrease in the dynamic range for the signal of interest caused by self-interference at the analog-to-digital interface. Using these system calculations, we provide comprehensive numerical results for typical transceiver parameters. The analytical results are also confirmed with full waveform simulations. We observe that the nonlinear distortion produced by the transmitter PA is a significant issue in a full-duplex transceiver and, when using cheaper and less linear components, also the receiver chain nonlinearities become considerable. It is also shown that, with digitally-intensive self-interference cancellation, the quantization noise of the ADCs is another significant problem.

\end{abstract}


\vspace{-2mm}
\section{Introduction}


\IEEEPARstart{F}{ull}-DUPLEX (FD) radio technology, where the devices transmit and receive signals simultaneously at the same center-frequency, is the new breakthrough in wireless communications. Such frequency-reuse strategy can theoretically double the spectral efficiency, compared to traditional half-duplex (HD) systems, namely time-division duplexing (TDD) and frequency-division duplexing (FDD). Furthermore, since the transmission and reception happen at the same time at the same frequency, the transceivers can sense each other's transmissions and react to them. This, with appropriate medium access control (MAC) design, can result in a low level of signaling and low latency in the networks. Because of these benefits, full-duplex radios can revolutionize the design of radio communications networks.

However, there are still several problems in the practical realization and implementation of small and low-cost full-duplex transceivers. The biggest challenge is the so called self-interference (SI), which results from the fact that the transmitter and receiver use either the same \cite{Knox12,Cox13} or separate but closely-spaced antennas \cite{Choi10,Jain11,Sahai11,Duarte10} and, thus, the transmit signal couples strongly to the receiver path. The power of the coupled signal can be, depending on, e.g., the antenna separation and transmit power, in the order of 60--100 dB stronger than the received signal of interest, especially when operating close to the sensitivity level of the receiver chain. In principle, the SI waveform can be perfectly regenerated at the receiver since the transmit data is known inside the device. Thus, again in principle, SI can be perfectly cancelled in the receiver path. However, because the SI signal propagates through an unknown coupling channel linking the transmitter (TX) and receiver (RX) paths, and is also affected by unknown nonlinear effects of the transceiver components, having perfect knowledge of the SI signal is, in practice, far from realistic.

In literature, some promising full-duplex radio demonstrations have recently been reported, e.g., in \cite{Choi10,Jain11,Sahai11,Duarte10}. In these papers, both radio frequency (RF) and digital signal processing (DSP) techniques were proposed for SI suppression. Nearly 70 to 80 dB of overall attenuation has been reported at best, but in real-world scenarios SI mitigation results have not been even nearly that efficient \cite{Jain11}. This is because with low-cost small-size electronics, feasible for mass-market products, the RF components are subject to many nonidealities compared to idealized demonstration setups reported in \cite{Choi10,Jain11,Sahai11,Duarte10}.

Several recent studies have analyzed selected analog/RF circuit non-idealities in the context of practical full-duplex radios. The phase noise of the transmitter and receiver oscillators has been analyzed, e.g., in \cite{Riihonen1222,Sahai12,Syrjala13,Ahmed132}. In these studies it was observed that the phase noise can potentially limit the amount of achievable SI suppression, especially when using two separate oscillators for transmitter and receiver. The effect of phase noise is also taken into account in the analysis presented in \cite{Sahai122}, where the feasibility of asynchronous full-duplex communications is studied. Furthermore, the impact of IQ mismatch induced mirror imaging has recently been addressd in \cite{Korpi133}.

\begin{figure*}[!t]
\centering
\includegraphics[width=\textwidth]{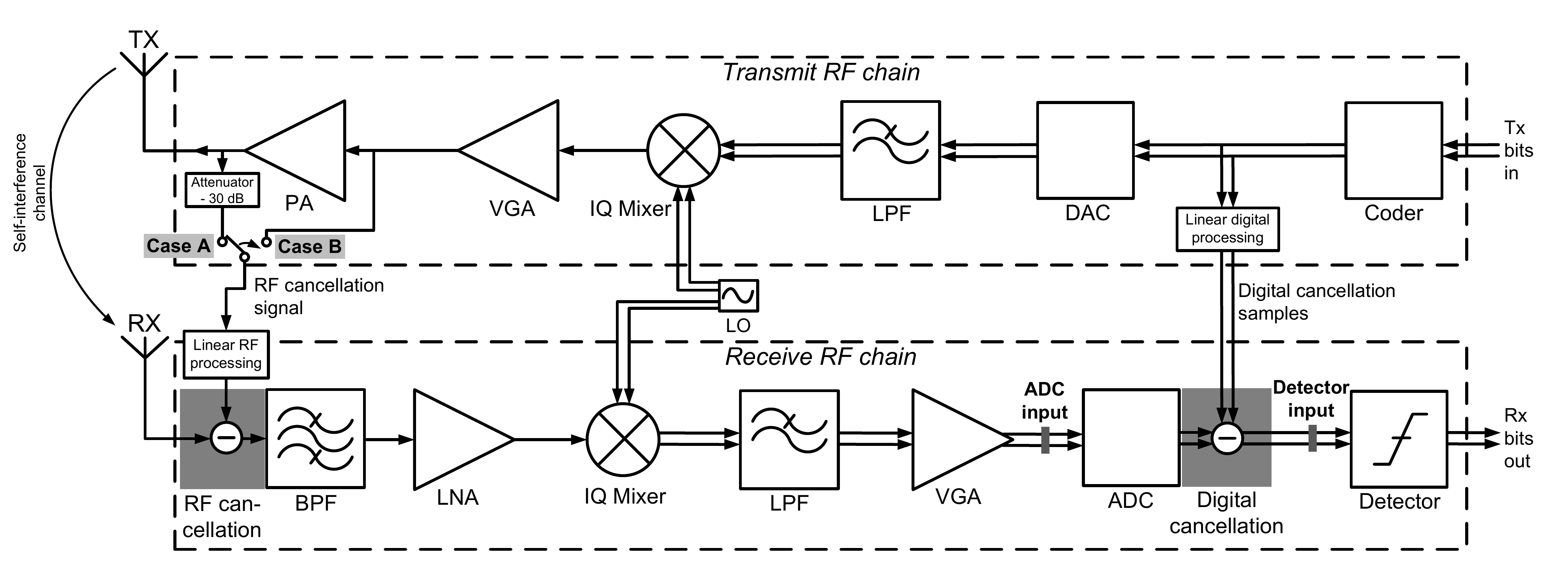}
\caption{Block diagram of the analyzed direct-conversion full-duplex transceiver, where RF and digital baseband interfaces for self-interference cancellation are illustrated in grey. Cases 'A' and 'B' refer to two alternative configurations assumed for reference signal extraction in RF cancellation.}
\label{fig:block_diagram}
\vspace{-4mm}
\end{figure*}

In addition, the amplifiers and mixers cause nonlinear distortion, especially with transmit powers in the order of 10--50 dBm that are typical for, e.g., mobile cellular radios. This can have a big impact on the characteristics and efficient cancellation of the SI waveform. Nonlinear distortion is a particularly important problem in full-duplex radios, since the receiver RF components must be able to tolerate the high-power SI signal, which is then gradually suppressed in the RX chain. Recently, the effect of nonlinear distortion in a full-duplex transceiver, and its compensation, have been studied, e.g., in \cite{Bharadia13,Ahmed13,Anttila13,Korpi132}. These studies indicate that nonlinear distortion of transceiver components, in particular with low-cost mass-product integrated circuits, forms a significant bottleneck in practical full-duplex radio devices.

Thus, in this paper, a comprehensive analysis of the nonlinear distortion effects in full-duplex transceivers is provided, with special focus on realistic achievable SI cancellation at receiver RF and DSP stages and corresponding maximum allowed transmit power. \emph{Such analysis and understanding is currently missing from the literature of the full-duplex field.} The analysis covers the effects of both transmitter and receiver nonlinearities, and shows that both can easily limit the maximum allowed transmit power of the device. Explicit expressions are provided that quantify the overall second- and third-order nonlinear distortion power, due to all essential RF components, at the detector input in the receiver. These can be used directly to, e.g., derive the required linearity figures for the transceiver RF components such that the nonlinear distortion at detector input is within any given implementation margin.

We also analyze, quantify, and compare two alternative RF cancellation strategies where reference signal is taken either from TX power amplifier (PA) input or output. We then show that PA nonlinearity can seriously limit the device operation already with transmit powers in the order of 5--10 dBm, especially when RF cancellation reference is taken from PA input. This indicates that, in addition to RX path, the linearity of the TX chain is also of high concern when designing and implementing full-duplex transceivers. The effect of transmit imperfections is also analyzed in \cite{Zheng13,Riihonen11,Day12,Day122} with a relatively simplified model. However, in this paper, the analysis of the transmit imperfections is done based on the actual properties of the TX components.

Finally, in addition to linearity analysis, the required dynamic range of the analog-to-digital converter (ADC) is addressed in this paper. Since a considerable amount of the SI cancellation is carried out in the digital domain, additional dynamic range is needed in the analog-to-digital interface or otherwise the SI signal heavily decreases the effective resolution of the weak desired signal. This, in turn, limits the performance of the whole transceiver. In this paper, we will explicitly quantify and derive the ADC dynamic range and resolution requirements such that the signal-to-interference-plus-noise ratio (SINR) at detector input in the RX chain does not degrade more than the specified implementation margin. \emph{Such analysis is also missing from the literature}. In particular, earlier work in \cite{Riihonen122} focuses on ADCs within an otherwise ideal system, while the current analysis incorporates the joint effect of quantization noise and all other nonidealities.

The organization of the rest of the paper is as follows. Section~\ref{sec:transceiver_model} describes the analyzed full-duplex direct-conversion transceiver model and especially the nonlinear characteristics of the essential TX and RX components. The system calculations, in terms of the powers of the useful and interfering signal components in different stages of the RX chain, as well as the required ADC performance, are then carried out and analyzed in Section~\ref{sec:results}. Section~\ref{sec:waveform_simul} provides the actual waveform-level reference simulation results of a complete full-duplex device, verifying the good accuracy of the system calculations and the associated performance limits. Finally, conclusions are drawn in Section~\ref{sec:conclusions}.

\emph{Nomenclature:} Throughout the paper, the use of linear power units is indicated by lowercase letters. Correspondingly, when referring to logarithmic power units, uppercase letters are used. The only exception to this is the noise factor, which is denoted by capital $F$ according to common convention in the literature of the field. Watts are used as the absolute power unit, and dBm as the logarithmic power unit.

\section{Full-duplex Transceiver Modeling}
\label{sec:transceiver_model}

\begin{figure*}[t]
        \centering
        \begin{subfigure}[t]{0.49\textwidth}
                \centering
                \includegraphics[width=\textwidth]{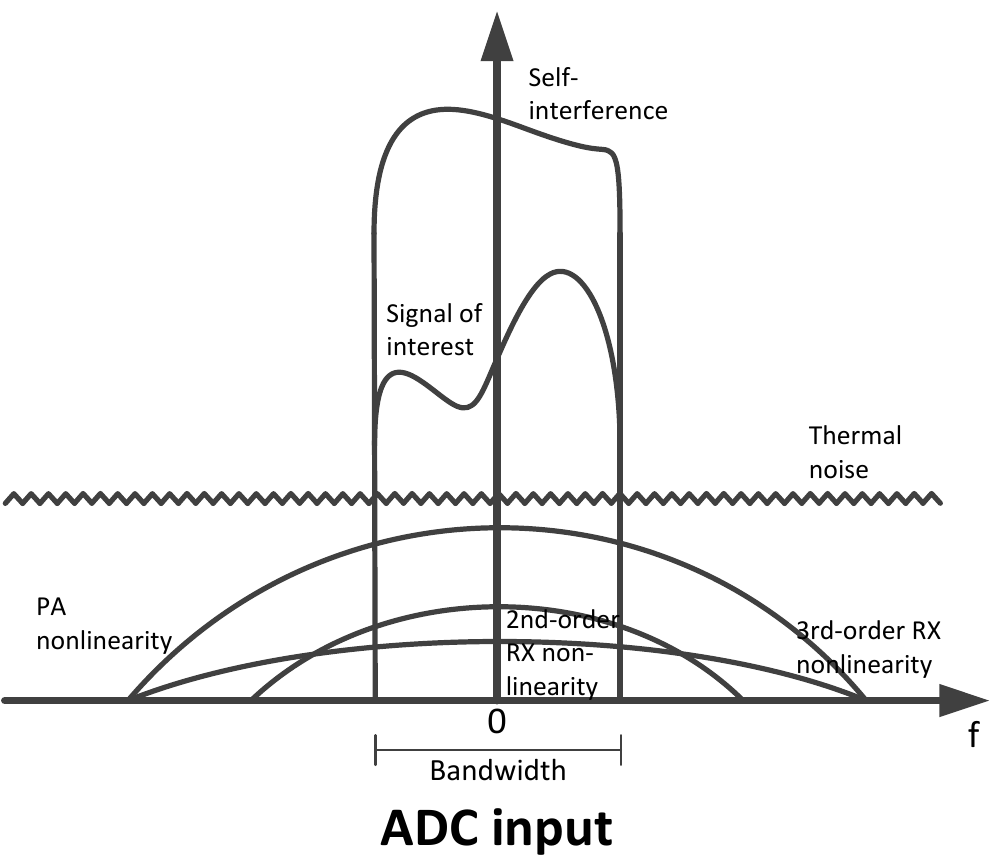}
                \caption{A sketch of the signal spectra at the ADC input.}
                \label{fig:spectra_adc}
        \end{subfigure}%
        ~
        \begin{subfigure}[t]{0.49\textwidth}
                \centering
                \includegraphics[width=\textwidth]{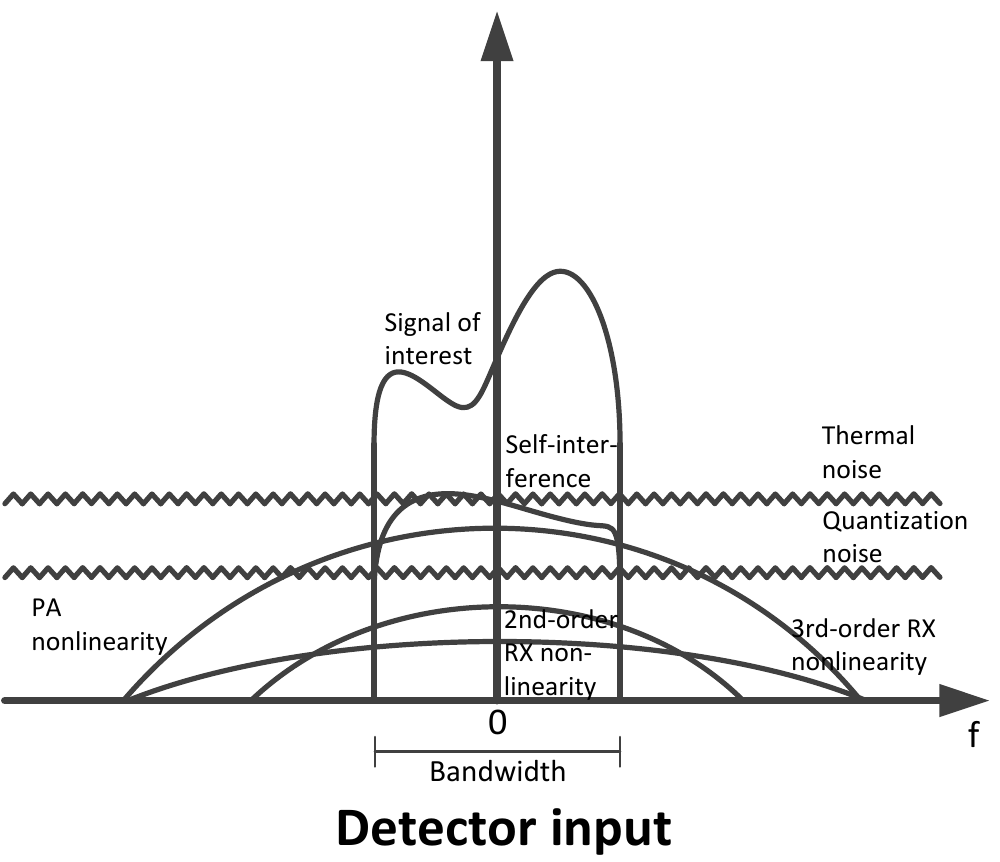}
                \caption{A sketch of the signal spectra at the detector input.}
                \label{fig:spectra_det}
        \end{subfigure}
        \label{fig:spectra}
        \caption{A principal illustration of the signal spectra at the inputs of the ADC and the detector. Note that this figure depicts a situation with a medium-level transmit power. With higher transmit powers, nonlinear distortion is more likely to be the dominant distortion component.}
\vspace{-4mm}
\end{figure*}

Our approach is to model a complete full-duplex transceiver component-wise, which allows us to analyze the feasibility of single-frequency full-duplex communications. Most of the emphasis in the calculations is at the receiver side since, due to the powerful self-interference, it is the more delicate part of the transceiver in terms of enabling full-duplex operation. Nevertheless, the effects of the transmitter are also taken into account as the exact SI waveform depends on, e.g., power amplifier nonlinearities. We wish to again emphasize that also oscillator phase noise can represent a performance bound in FD devices \cite{Riihonen1222,Sahai12,Syrjala13,Ahmed132}. However, as the focus in this article is on nonlinear distortion and ADC interface, phase noise is neglected in the following.

A block diagram representing the analyzed full-duplex direct-conversion transceiver is given in Fig.~\ref{fig:block_diagram}. For generality, both RF- and DSP-based SI cancellation \cite{Riihonen122} are covered in the analysis. The direct-conversion architecture is chosen due to its simple structure and wide applications, e.g., in cellular devices. Another signifcant aspect is the assumed reference signal path for RF cancellation. In this paper, two alternative scenarios are analyzed: Case A, in which the reference signal is taken from the output of the PA and attenuated to a proper level, and Case B, in which the reference signal is taken directly from the input of the PA. These scenarios are also marked in the block diagram in Fig.~\ref{fig:block_diagram} using a switch. In general, both RF and DSP cancellation stages are assumed to deploy only linear processing.

\subsection{Analysis Principles and Performance Measures}
\label{sec:principles_and_measures}

In the transceiver system calculations, the two most relevant interfaces are the \emph{ADC input} and \emph{detector input}. These points are also marked in the block diagram in Fig.~\ref{fig:block_diagram}. Furthermore, example signal characteristics and the different signal components, alongside with their typical relative power levels, are illustrated in Figs.~\ref{fig:spectra_adc} and \ref{fig:spectra_det}. The reason for the significance of the ADC input is the role of quantization and its dependence on SI. As the receiver automatic gain control (AGC) keeps the total ADC input at constant level, higher SI power means reduced desired signal power and thus more and more of the ADC dynamic range is reserved by the SI signal. This, in turn, indicates reduced effective resolution for the desired signal, which may limit the receiver performance.

The effect of quantization is studied by determining the SINR at the ADC input, quantifying the power of the desired signal relative to the other signal and distortion components at this point. A typical situation in terms of the power levels at this interface can be seen in Fig.~\ref{fig:spectra_adc}, where the SI signal is clearly dominating, and thus reserving a signicant amount of dynamic range.

Then, to characterize the overall performance of the whole full-duplex transceiver, and how the different types of distortion affect it, also the final SINR at the detector input, including digital SI cancellation, is studied and analyzed. This is thus the other significant point or calculation interface in the forthcoming analysis. Typical power levels also at this interface can be seen in Fig.~\ref{fig:spectra_det}, where the SI signal has now been attenuated by digital cancellation, and it is no more the dominant distortion component. However, due to analog-to-digital conversion, there is now quantization noise in the total signal, which might be a significant issue, depending on the parameters of the transceiver.

Throughout the rest of the article, it is assumed that all the distortion types can be modelled in additive form. This is very typical in transceiver system calculations, see, e.g., \cite{Gu06,Razavi98}. The good accuracy of this approach is also verified by full waveform simulations later in Section~\ref{sec:waveform_simul}. 

Under the above assumptions, the SINR on linear scale at the ADC input can now be directly defined as
\begin{align} \label{eq:sinr_adc}
\mathit{sinr}_{\text{ADC}} = \frac{g_\text{rx}p_{\text{SOI,in}}}{g_\text{rx}Fp_{\text{N,in}} + \frac{g_\text{rx}}{a_{\text{ant}}}\left( \frac{p_{\text{tx}}}{a_{\text{RF}}} + \frac{p_{\text{3rd,PA,tx}}}{a_{\text{NL}}} \right) + p_{\text{2nd}} + p_{\text{3rd}}} \text{,}
\end{align}
where $ g_\text{rx} $ is the total gain of the RX chain, $ p_{\text{SOI,in}} $ is the power of the signal of interest at RX input, $ F $ is the noise factor of the receiver, $ p_{\text{N,in}} $ is the thermal noise power at the input of the receiver, $ a_{\text{ant}} $ and $ a_{\text{RF}} $ are the amounts of antenna attenuation and RF cancellation, $ p_{\text{tx}} $ is the transmit power, $p_{\text{3rd,PA,tx}}$ is the power of PA-induced nonlinear distortion at the output of the transmit chain, parameter $a_{\text{NL}}$ is $a_{\text{RF}}$ for Case A and $1$ for Case B, and $ p_{\text{2nd}} $ and $ p_{\text{3rd}} $ are the cumulated powers of 2nd- and 3rd-order nonlinear distortion produced at the RX chain. All the powers are assumed to be in linear units, which is indicated also by the lowercase symbols. These signal components are illustrated in Fig.~\ref{fig:spectra_adc} with realistic relative power levels.

The purpose of defining the ADC input SINR is to quantify the ratio of the useful signal power and total noise-plus-interference power entering the analog-to-digital interface. With fixed ADC voltage range, and assuming that the overall receiver gain is controlled properly, the total ADC input power $g_\text{rx}p_\text{SOI,in} + g_\text{rx}Fp_{\text{N,in}} + \frac{g_\text{rx}}{a_{\text{ant}} a_{\text{RF}}}p_{\text{tx}} + \frac{g_\text{rx}}{a_{\text{NL}}}p_{\text{3rd,PA,tx}} + p_{\text{2nd}} + p_{\text{3rd}}$ is always matched to the maximum allowed average power, say $p_\text{target}$. This will be elaborated in more details later.

Taking next the quantization noise and digital cancellation into account, the SINR at the detector input can be defined as
\begin{align} \label{eq:sinr_det}
\mathit{sinr}_{\text{D}} = \frac{g_\text{rx}p_{\text{SOI,in}}}{\splitfrac{g_\text{rx}Fp_{\text{N,in}} + \frac{g_\text{rx}}{a_{\text{ant}}}\left( \frac{p_{\text{tx}}}{a_{\text{RF}} a_\text{dig}} + \frac{p_{\text{3rd,PA,tx}}}{a_{\text{NL}}} \right) + p_{\text{quant}}+\cdots} { + p_{\text{2nd}} + p_{\text{3rd}}}} \text{,}
\end{align}
where $ a_{\text{dig}} $ is the attenuation achieved by digital cancellation and $ p_{\text{quant}} $ is the power of quantization noise. This SINR defines the overall receiver performance of the full-duplex transceiver and is thus the most significant figure of merit in the analysis. A realistic scetch of the relative power levels of the specified signal components also at this interface can be seen in Fig.~\ref{fig:spectra_det}.

The following subsections analyze in detail the different component powers of the above two principal equations, and their dependence on the transmit power, RF cancellation, digital cancellation, and TX and RX chain nonlinear characteristics. Then, in Section~\ref{sec:results}, these are all brought together and it is analyzed in detail how these elementary parameters and transceiver characteristics affect the SINR at both of the studied interfaces and thereon the whole transceiver operation.

\subsection{Radio-Frequency Front-End}
\label{sec:receiver}
\subsubsection{Receiver Reference Sensitivity}

The most challenging situation from the SI suppression perspective is when the actual received signal is close to the receiver sensitivity level. Thus, we begin by briefly defining the receiver reference sensitivity, which is determined by the thermal noise floor at RX input, the noise figure of the receiver, and the signal-to-noise ratio (SNR) requirement at the detector. This forms then the natural reference for assumed received signal levels in our analysis. The reference sensitivity, expressed in dBm, follows directly from \cite{Gu06} and can be written as
\begin{align} \label{eq:sens2}
P_{\text{sens}} = -174+10\log_{10}(B)+\mathit{NF}_{\text{rx}}+\mathit{SNR}_{\text{d}} \text{,}
\end{align}
where $B$ is the bandwidth of the system in Hertz, $\mathit{NF}_{\text{rx}}$ is the noise figure of the receiver, and $\mathit{SNR}_{\text{d}}$ is the SNR requirement at the input of the detector. In modern radio systems, the sensitivity is, strictly-speaking, affected by the assumed code rate and modulation through varying SNR requirements. However, for simplicity, only two reference sensitivity numbers are assumed in this study in the numerical examples of Section~\ref{sec:results}.

The total receiver noise figure, in dB, is in general defined as $\mathit{NF}_{\text{rx}} = 10\log_{10}(F_{\text{rx}})$ where the total noise factor of the assumed RX chain in Fig.~\ref{fig:block_diagram} is given by the classical Friis' formula \cite{Gu06} as
\begin{align} \label{eq:nfac}
F_{rx} = F_{\text{LNA}}+\frac{F_{\text{mixer}}-1}{g_{\text{LNA}}}+\frac{F_{\text{VGA}}-1}{g_{\text{LNA}}g_{\text{mixer}}} \text{.}
\end{align}
\noindent In above, $F_{\text{LNA}}$, $F_{\text{mixer}}$, and $F_{\text{VGA}}$ are the noise factors of the LNA, IQ Mixer, and VGA, respectively. Similarly, $g_{\text{LNA}}$, $g_{\text{mixer}}$, and $g_{\text{VGA}}$ are the linear gains of the components.

\subsubsection{RF Cancellation}

In general, depending on the antenna separation, the path loss between the transmit and receive antennas attenuates the SI signal to a certain degree. However, to prevent the saturation of the RX chain, additional RF cancellation is most likely required. For generality, a multi-tap RF cancellation circuit, as presented in \cite{Bharadia13,Choi13}, is assumed in this paper. This type of a cancellation circuit consists of several fixed delay lines, each of which has its own weight factor. Thus, the final cancellation signal consists of a linear combination of several delayed versions of the reference transmit signal with appropriate phase and amplitude tuning. The cancellation is then done by estimating the coefficients for the different delay lines based on the SI coupling channel, and subtracting this cancellation signal from the received signal. Thus, depending on the chosen delays, this type of a RF cancellation scheme might even be able to attenuate the multipath components. It should be noted, however, that in terms of the actual system calculations there is no difference between using a single- or multi-tap RF canceller, as only the amount of achieved SI attenuation is taken into account by the equations.

Furthermore, in our analysis, two alternatives for the reference signal path are considered, as follows.
\begin{itemize}
\item Case A describes perhaps the most widely used implementation technique for taking the reference signal for RF cancellation \cite{Jain11,Choi10,Shenghong11,Radunovic09,Chen98}. However, the drawback of this approach is the need for a bulky RF attenuator to achieve a feasible power level for the cancellation signal. The required amount of attenuation is obviously the estimated path loss between the antennas, as this ensures that the powers of the reference signal and incoming SI signal are of similar magnitude at the RF cancellation block.
\item In Case B, the reference signal is taken already from the input of the PA. As the gain of the PA is usually within 10 dB of the magnitude of the path loss between the antennas \cite{Gu06,Razavi98}, only a tunable amplitude and phase matching circuit, such as a commercial product \cite{qhx220}, with feasible tuning range is required. Thus, no additional RF attenuator is needed, resulting in a simpler and lower-cost RF frontend. On the other hand, as shown in this paper, the problem in this implementation is the nonlinear distortion produced by the PA, which is not included in the reference signal. Thus, it is not attenuated by RF cancellation like in Case A, resulting in a lower SINR in the analog domain. This will be illustrated in Section~\ref{sec:results}. Notice also that from the PA nonlinearity perspective, Case B is equivalent to the method used in \cite{Sahai11} and \cite{Duarte12}, where a separate low-power TX chain is used to generate the RF reference signal. Thus, in our analysis, Case B covers indirectly also this type of transceiver scenarios.
\end{itemize}

\subsection{Analog-to-Digital Interface and Digital Cancellation}

Next we address issues related to analog-to-digital interface and quantization noise, especially from the perspective of residual SI left for digital cancellation. The starting point is the classical expression, available in, e.g., \cite{Gu06}, defining the signal-to-quantization-noise ratio of the ADC as
\begin{align} 
\mathit{SNR}_{\text{ADC}} &= 6.02 b + 4.76 - \mathit{PAPR} \text{,} \label{eq:adc_snr}
\end{align}
where $b$ is the number of bits at the ADC, and $\mathit{PAPR}$ is the estimated peak-to-average power ratio. The above expression assumes proper AGC at ADC input such that the full range of the ADC is used but the clipping of the signal peaks is avoided. However, the analysis could be easily translated to cover clipping noise as well \cite{Riihonen122}. 

Building on the above expression, our approach to analyze the impact of SI on analog-to-digital interface is to determine how many bits are effectively lost from the signal of interest. This is directly based on the fact that the remaining SI signal reserves part of the dynamic range of the ADC and thus decreases the resolution of the desired signal. Now, the amount of lost bits due to RX noise and interference can be determined by calculating how many dBs the signal of interest is below the total signal power, as this is directly the amount of dynamic range that is reserved by the noise and interference. The amount of lost bits can thus in general be calculated from \eqref{eq:adc_snr} as
\begin{align}
	b_\text{lost,I+N} = \frac{P_\text{tot}-P_\text{SOI}}{6.02} \text{,} \label{eq:bitloss_basic}
\end{align}
where $P_\text{tot}$ and $P_\text{SOI}$ are the total power of the signal and the power of the signal of interest at the input of the ADC, respectively, and 6.02 depicts the dynamic range of one bit, thus mapping the loss of dynamic range to loss of bits. Then, the actual \emph{bit loss due to self-interference} is defined as the increase in lost bits when comparing the receiver operation with and without SI. Following this step-by-step path, and using \eqref{eq:bitloss_basic}, a closed-form equation for the bit loss can be derived as shown in detail in Appendix~\ref{app:bitloss}, yielding
\begin{align}\label{eq:bitloss}
b_{\text{lost}} = \log_{4} & \left[1 +\left( \frac{1}{p_\text{SOI,in} + p_\text{N,in}}\right) \right. \nonumber\\
&\left. {} \cdot \left(\frac{p_\text{tx}}{a_\text{ant} a_\text{RF}} + \frac{p_\text{tx}^3}{a_\text{ant} a_\text{NL} \mathit{iip3}_\text{PA}^2 g_\text{PA}^2}\right)\right] \text{.}
\end{align}
{Here, $\mathit{iip3}_{\text{PA}}$ and $g_{\text{PA}}$ are the IIP3 figure and gain of the PA in linear units, respectively}. 

An immediate observation following from (\ref{eq:bitloss}) is that increasing the transmit power with respect to the other signal components also increases the bit loss. Furthermore, increasing antenna attenuation or RF cancellation decreases the bit loss. These are relatively intuitive results, but with (\ref{eq:bitloss}) they can be quantified and analyzed exactly. It is also important to note that the bit loss does not depend on the total amount of bits in the ADC. Thus, the detailed numerical illustrations given in Section~\ref{sec:results}, based on (\ref{eq:bitloss}), apply to all ADCs.

Finally, prior to detection, the remaining SI is mitigated in the digital domain by subtracting the transmitted baseband waveform from the received signal. The subtracted samples are generated by linearly filtering the transmitted symbols with an estimate of the overall coupling channel response linking the TX and RX. In practice, the channel estimation at this stage includes the effects of the transmitter, the coupling channel between the antennas, and the receiver. Also the multipath components due to reflections are included in the channel estimate. In our analysis, as was already illustrated in (2), the efficiency of digital cancellation is parameterized through digital SI attenuation $a_\text{dig}$, or $A_\text{dig}$ in dB. Notice that since only linear digital cancellation is assumed, only the linear SI component is suppressed.

\subsection{Nonlinear Distortion in Receiver Chain}
\label{sec:nlDist}

In addition to quantization noise, the nonlinear distortion produced by the components of the transceiver is also of great interest. Following the well-established conventions from literature, nonlinear distortion of individual components is modeled by using the IIP2 and IIP3 figures (2nd- and 3rd-order input-referred intercept points) \cite{Gu06}. For a general $n$th-order nonlinearity, the power of the nonlinear distortion in dBm at the output of the component is given by
\begin{align} \label{eq:nonlinear}
P_{\text{nth}} = P_{\text{out}} - (n-1)(\mathit{IIPn} - P_{\text{in}}) \text{,}
\end{align}
where $P_{\text{in}}$ is the total input power of the component, $P_{\text{out}}$ is the total output power, and $IIPn$ is $n$th-order input-referred intercept point, all in dBm. As is well known in the literature, such principal power characteristics apply quite accurately, given that the component is not driven to full saturation, while offering analytically tractable expressions to accumulate total nonlinear distortion powers of a complete transceiver chain. In the case of the RX chain, this includes the LNA, mixers, and baseband VGA. The accuracy of this approach over a wide range of parameters, e.g., transmit powers, is illustrated and verified through full reference waveform simulations in Section~\ref{sec:waveform_simul}.

\subsection{Transmitter Modeling and PA Nonlinearity}
\label{sec:transmitter}

When analyzing and modeling the TX chain, it is assumed that the power of thermal noise is negligibly low. This is a reasonable assumption as transmitters are never limited by inband thermal noise floor. Hence, thermal noise is omitted from transmitter modeling and only injected at RX input. Furthermore, we also assume that the power amplifier is the main source of nonlinear distortion, since all other transmitter components operate at low power regime. In fact, even if some nonlinear distortion was created, e.g., in the feeding amplifier prior to PA, it is a part of the RF cancellation reference signal in all the considered scenarios, and hence suppressed by RF cancellation below the RX noise level. Thus, it is sufficient to focus on the nonlinearities of the PA when analyzing the transmitter.

The PA itself, in turn, is typically heavily nonlinear \cite{Gu06,Razavi98,Raab02}. In our analysis, we assume that the PA produces 3rd-order distortion which falls on to the signal band, since this is the dominant distortion in practice. This is characterized with the IIP3 figure of the PA, according to \eqref{eq:nonlinear}. Furthermore, in Case~A, this distortion is included in the reference signal, and is thus attenuated by RF cancellation. In Case~B, this is not the case, and the nonlinear distortion produced by the PA remains at the same level after RF cancellation, as it is only attenuated by the coupling channel path loss.

Another observation about the nonlinearities of the transmit chain is that linear digital cancellation cannot suppress them{, because} the reference symbols for digital cancellation exist only in the digital domain and do not include any analog distortion. Moreover, nonlinear distortion cannot be modelled with a linear filter, and thus linear digital cancellation is unable to mitigate it. \emph{The results shown in this paper thus give motivation to develop nonlinear digital SI cancellation techniques}. First works to this direction have been very recently reported in \cite{Anttila13,Bharadia13,Ahmed13}.

\subsection{Accumulated Component Powers at Detector Input}
\label{sec:equations}

The previous subsections describe elementary component-level modeling principles. Next, in this subsection, we accumulate the total observable power levels of all essential individual signal components at the \emph{input of the detector}. This includes the desired signal power, (residual) SI power, quantization noise power, thermal noise power, RX 2nd- and 3rd-order nonlinear distortion power, and TX PA induced 3rd-order nonlinear distortion power. 

First, the power of the quantization noise at the detector input can be written as
\begin{align}
P_{\text{quant}} = P_{\text{target}}-\mathit{SNR}_{\text{ADC}} = P_{\text{target}} - 6.02 b - 4.76 + \mathit{PAPR} \text{,} \label{eq:p_quant}
\end{align}
where $P_\text{target}$ is the maximum allowed average power of the signal at the ADC input, such that clipping is avoided. For any given PAPR, it can be observed that the power of the quantization noise depends only on the characteristics of the ADC, namely its maximum input power and the amount of bits.

The powers of the other signal components depend on several parameters, first and foremost on the total gain of the RX chain. As the signal of interest, SI signal, and the nonlinear distortion produced by the PA are the only significant signal components at the very input of the receiver, the total gain in linear units can be first written as
\begin{align} \label{eq:gain_real}
g_\text{rx} = \frac{p_{\text{target}}}{\frac{1}{a_{\text{ant}}} \left( \frac{p_{\text{tx}}}{ a_{\text{RF}}}+\frac{p_{\text{3rd,PA,tx}}}{a_{\text{NL}}}\right) + p_{\text{SOI,in}}} \text{.}
\end{align}
When considering Case A, the nonlinear distortion produced by the PA is attenuated by RF cancellation. Thus, with high transmit powers, the power of the total signal at the input of the receiver can be approximated by the power of SI, as it is several orders of  magnitude higher than the power of any other signal component when operating close to the sensitivity level. In this case, \eqref{eq:gain_real} is simplified to
\begin{align}
g_\text{rx} = \frac{a_{\text{ant}} a_{\text{RF}} p_{\text{target}}}{p_{\text{tx}}} \text{.} \label{eq:gain_appr}
\end{align}

Knowing now the total gain of the receiver, it is then trivial to write the expressions for the powers of the other signal components, namely the signal of interest and thermal noise, at the input of the detector in dBm as
\begin{align} 
P_{\text{SOI}} = P_{\text{SOI,in}}+G_\text{rx} \text{ and} \label{eq:p_soi}
\end{align}
\begin{align}
P_{\text{N}} = P_{\text{N,in}}+G_\text{rx}+\mathit{NF}_{\text{rx}} \text{.} \label{eq:p_n}
\end{align}
The corresponding power of linear SI can be written as
\begin{align}
P_{\text{SI}} = P_{\text{tx}}-A_{\text{ant}}-A_{\text{RF}}-A_{\text{dig}}+G_\text{rx} \text{.} \label{eq:p_si}
\end{align}
Furthermore, for high transmit powers, when (\ref{eq:gain_appr}) can be used to approximate the total gain of the RX chain, \eqref{eq:p_si} becomes $ P_{\text{SI}} = P_{\text{target}}-A_{\text{dig}} $.

Next, the total powers of the 2nd- and 3rd-order nonlinear distortion, produced by the RX chain, are derived based on \eqref{eq:nonlinear}, as shown in detail in Appendix~\ref{app:nonlinear}. The resulting equations are
\begin{align} \label{eq:p2}
p_{\text{2nd}} \approx g_{\text{LNA}}^2 g_{\text{mixer}} g_{\text{VGA}} p_{\text{in}}^2 \left(\frac{1}{\mathit{iip2}_{\text{mixer}}}+\frac{g_{\text{mixer}}}{\mathit{iip2}_{\text{VGA}}} \right)
\end{align}
\begin{align}
p_{\text{3rd}} &\approx g_{\text{LNA}} g_{\text{mixer}} g_{\text{VGA}} p_{\text{in}}^3 \left[\left(\frac{1}{\mathit{iip3}_{\text{LNA}}}\right)^2\right.\nonumber\\
&\left. {}+\left(\frac{g_{\text{LNA}}}{\mathit{iip3}_{\text{mixer}}}\right)^2+\left(\frac{g_{\text{LNA}} g_{\text{mixer}}}{\mathit{iip3}_{\text{VGA}}}\right)^2 \right] \text{,} \label{eq:p3}
\end{align}
where the subscript of each parameter indicates the considered component. Furthermore, $\mathit{iip2}_{\text{k}}$ and  $\mathit{iip3}_{\text{k}}$ are the 2nd- and 3rd-order input intercept points expressed in Watts, $g_{\text{k}}$ is the linear gain of the corresponding component, and $ p_{\text{in}} $ is the total power of the signal after RF cancellation, again in Watts.

Finally, the power of the PA-induced nonlinear distortion at the output of the transmit chain can be written as
\begin{align} 
P_{\text{3rd,PA,tx}} &= P_{\text{tx}}-2(\mathit{IIP3}_{\text{PA}}-(P_{\text{tx}}-G_{\text{PA}}))\nonumber\\
&=3P_{\text{tx}}-2(\mathit{IIP3}_{\text{PA}}+G_{\text{PA}})\text{,} \label{eq:p3_pa_tx}
\end{align}
This value is used in, for example, \eqref{eq:gain_real}, as the gain is determined based on the signal levels at the input of the RX chain. The power of the PA-induced nonlinear distortion at the input of the detector can then be written as
\begin{align} 
P_{\text{3rd,PA}} &= P_{\text{3rd,PA,tx}} + G_\text{rx} - A_\text{ant} - A_\text{NL}\nonumber\\
&=3P_{\text{tx}}-2(\mathit{IIP3}_{\text{PA}}+G_{\text{PA}}) + G_\text{rx} - A_\text{ant} - A_\text{NL}  \text{.} \label{eq:p3_pa}
\end{align}
As only linear digital cancellation is deployed, the nonlinear distortion produced by the PA is only attenuated by the coupling channel path loss ($A_\text{ant}$), and potentially by RF cancellation ($A_\text{NL} = A_\text{RF}$), if considering Case A. Further attenuation of this nonlinear component with actual nonlinear cancellation processing, analog or digital, is out of the scope of this paper. The potential benefits of digitally attenuating nonlinearly distorted SI signals are analyzed in, e.g., \cite{Anttila13,Ahmed13,Bharadia13}.

\section{System Calculations and Results}
\label{sec:results}

In this section, we put together the elementary results of the previous section in terms of overall system calculations. The basic assumption is that the actual received signal is only slightly above the receiver sensitivity level, as this is the most challenging case from the SI perspective. The main interests of these calculations are then to see how much the quantization noise produced by the ADC affects the overall performance of the transceiver, and how severe the nonlinear distortion products, caused by full-duplex operation, are at the detector input. For this reason, the final signal quality ($\mathit{SINR}_{\text{d}}$) after the ADC and digital cancellation is measured with different parameters and transmit powers.

In all the experiments, the maximum allowed SINR loss due to full-duplex operation is assumed to be 3 dB. This means that if the effective total noise-plus-interference power more than doubles compared to classical half-duplex operation, then the receiver performance loss becomes too high. Thus, the derived $\mathit{SINR}_{\text{d}}$ values under FD operation are compared to signal-to-thermal-noise-ratio ($\mathit{SNR}_{\text{d}}$) at the input of the detector. The transmit power level at which this 3 dB loss is reached is referred to as the \textit{maximum transmit power}. It is marked to all relevant result figures with a vertical line to illustrate what is effectively the highest transmit power with which the full-duplex transceiver can still operate with tolerable SINR loss.

This also provides a way to obtain some insight into the relative performances of half-duplex and full-duplex radio devices. Namely, with a low SINR loss, a full-duplex radio can be assumed to approximately double the spectral efficiency, whereas with a high SINR loss, the effective spectral efficiency might be even lower than that achieved by traditional half-duplex radios. A SINR loss of 3 dB illustrates a point at which full-duplex transceivers can still be expected to provide a capacity gain in comparison to half-duplex transceivers. In addition, \eqref{eq:bitloss} compares the effect of quantization noise in full-duplex and half-duplex transceivers by determining the SI-induced decrease in the effective dynamic range of the ADC. However, a more in-depth analysis regarding the performance of practical full-duplex transceivers, especially at system/network level, is out of the scope of this paper, and we consider it as a possible topic for future work. Furthermore, a detailed performance comparison between half-duplex and full-duplex radios under some implementation impairments, excluding nonlinear distortion, is already done in \cite{Ahmed133}.

\subsection{Parameters for Numerical Results}
\label{sec:parameters}

In order to provide actual numerical results with the derived equations, parameters for the full-duplex transceiver are specified. It should be emphasized that the chosen parameters are just example numbers chosen for illustration purposes only, and all the calculations can be easily repeated with any given parametrization.

\subsubsection{Receiver}
\label{sec:calc_receiver}

The general system level parameters of the studied full-duplex transceiver are shown in Table~\ref{table:system_parameters}, and the parameters of the individual components of the receiver are shown in Table~\ref{table:rx_parameters}. Two sets of parameters are used, which are referred to as Parameter Set~1 and Parameter Set~2. The first set of parameters corresponds to state-of-the-art wideband RF transceiver performance. The parameters of the 2nd set model a more challenging scenario with lower received signal power, decreased linearity, and slightly inferior SI cancellation ability. In most parts of the analysis, Parameter Set~1 is used as it depicts better the characteristics of modern transceivers, especially in terms of bandwidth and linearity.

With (\ref{eq:sens2}), the sensitivity level of the receiver can be calculated as $P_{\text{sens}} = -88.9 \text{ dBm}$ for Parameter Set~1. This is a typical realistic value and close to the reference sensitivity specified in the LTE specifications \cite{LTE_specs}. For Parameter Set~2, the corresponding sensitivity is $P_{\text{sens}} = -100.1 \text{ dBm}$, which is an even more challenging value{, assuming that the power of the received signal is close to the sensitivity level}. Here, the power of the received signal is assumed to be 5 dB above sensitivity level, resulting in a received power level of either $P_{\text{SOI,in}} = -83.9 \text{ dBm}$ or $P_{\text{SOI,in}} = -95.1 \text{ dBm}$, depending on the parameter set.

\begin{table}[!t]
\renewcommand{\arraystretch}{1.3}
\caption{System level parameters of the full-duplex transceiver for Parameter Sets 1 and 2.}
\label{table:system_parameters}
\centering
\begin{tabular}{|c||c||c|}
\hline
\textbf{Parameter} & \textbf{Value for } & \textbf{Value for } \\
& \textbf{Param. Set 1} & \textbf{Param. Set 2}\\
\hline
SNR requirement & 10 dB & 5 dB\\
\hline
Bandwidth & 12.5 MHz & 3 MHz\\
\hline
Receiver noise figure & 4.1 dB & 4.1 dB\\
\hline
Sensitivity & $-$88.9 dBm & $-$100.1 dBm\\
\hline
Received signal power & $-$83.9 dBm & $-$95.1 dBm\\
\hline
Antenna separation & 40 dB & 40 dB\\
\hline
RF cancellation & 40 dB & 20 dB\\
\hline
Digital cancellation & 35 dB & 35 dB\\
\hline
ADC bits & 8 & 12\\
\hline
ADC P-P voltage range & 4.5 V & 4.5 V\\
\hline
PAPR & 10 dB & 10 dB\\
\hline
Allowed SINR loss & 3 dB & 3 dB\\
\hline
\end{tabular}
\end{table}

\begin{table}[!t]
\renewcommand{\arraystretch}{1.3}
\caption{Parameters for the components of the receiver. The values in the parentheses are the values used in Parameter Set 2.}
\label{table:rx_parameters}
\centering
\begin{tabular}{|c||c||c||c||c|}
\hline
\textbf{Component} & \textbf{Gain [dB]} & \textbf{IIP2 [dBm]} & \textbf{IIP3 [dBm]} & \textbf{NF [dB]}\\
\hline
BPF & 0 & - & - & 0\\
\hline
LNA & 25 & 43 & $-$9 ($-$15) & 4.1\\
\hline
Mixer & 6 & 42 & 15 & 4\\
\hline
LPF & 0 & - & - & 0\\
\hline
VGA & 0--69 & 43 & 14 (10) & 4\\
\hline
Total & 31--100 & 11 & $-$17 ($-$21) & 4.1 \\
\hline
\end{tabular}
\end{table}

The isolation between the antennas is assumed to be 40 dB. This value, or other values of similar magnitude, have been reported several times in literature \cite{Jain11, Sahai11, Duarte10}. Furthermore, the assumed RF cancellation level for Parameter Set~1 is 40 dB. For a single-tap RF canceller (used, e.g., in \cite{Choi10,Jain11}), this value is somewhat optimistic However, if a multi-tap RF cancellation circuit is considered, RF cancellation values of this magnitude can easily be expected \cite{Bharadia13}. In Parameter Set~2, in turn, a lower RF cancellation level of 20 dB is assumed to represent a more practical scenario.

The component parameters of the actual direct-conversion RX chain are determined according to \cite{Yoshida03,Parssinen99,Behzad07}. The objective is to select typical parameters for each component, and thus obtain reliable and feasible results. The chosen parameters are shown in Table~\ref{table:rx_parameters}, where the values without parentheses are used with Parameter Set 1, while the values with parentheses are used with Parameter Set 2. With (\ref{eq:p2}) and (\ref{eq:p3}), the total IIP2 and IIP3 figures of the whole receiver can be calculated to be $10.8$ dBm and $-17.1$ dBm (Parameter Set 1) or $10.8$ dBm and $-20.1$ dBm (Parameter Set 2), respectively.

The ADC input is controlled by the VGA such that the assumed full voltage range is properly utilized. As a realistic scenario, PAPR of the total signal is assumed to be 10 dB and state-of-the-art ADC specifications in \cite{ADC_datasheet} are deployed in terms of full voltage range. Using now (\ref{eq:adc_snr}), the signal-to-quantization noise ratio of the ADC is $ \mathit{SNR}_{\text{ADC}} = 6.02b -5.24 $, where $b$ is the number of bits at the ADC.

\subsubsection{Transmitter}
\label{sec:calc_transmitter}

The parameters of the individual TX components are shown in Table~\ref{table:tx_parameters}{, and they are the same for both parameter sets.} Again, typical values are chosen for the parameters according to \cite{Gu06} and \cite{Razavi98}. This ensures that the conclusions apply to a realistic TX chain. Furthermore, for the transmitter, only 3rd-order nonlinear distortion is taken into account as the 2nd-order nonlinearities do not fall on the actual signal band. Assuming that the power of the feeding amplifier input signal is approximately $-$35~dBm, it can be observed from the table that, with the maximum feeding amplifier gain, the power of the 3rd-order nonlinear distortion at the output of the transmit chain is 40 dB lower than the fundamental signal component. Hence, the spectral purity of the considered TX chain is relatively high, and thus the obtained results, when it comes to the PA-induced nonlinear distortion, are on the optimistic side.

\begin{table}[!t]
\renewcommand{\arraystretch}{1.3}
\caption{Parameters for the components of the transmitter.}
\label{table:tx_parameters}
\centering
\begin{tabular}{|c||c||c||c|}
\hline
\textbf{Component} & \textbf{Gain [dB]} & \textbf{IIP3 [dBm]} & \textbf{NF [dB]}\\
\hline
LPF & 0  & - & 0 \\
\hline
Mixer & 5 & 5 & 9\\
\hline
VGA & 0--35 & 5 & 10\\
\hline
PA & 27 &  20 & 5\\
\hline
Total & 32--67 & $-$20 & 10.3\\
\hline
\end{tabular}
\end{table}

Taking into account the input power and maximum gain range of the feeding amplifier, it can also be observed From Table~\ref{table:tx_parameters} that the power of the transmitted signal is between $-$8 and 27 dBm. This is a sufficient range for example in WLAN applications, or in other types of indoor communications. In addition, the studied transmit power range applies in some cases also to mobile devices in a cellular network, like class 3 LTE mobile transmitter \cite{LTE_specs}. In the following numerical results, the transmit power is varied between $-$5 and 25 dBm.

\subsection{Results with Case A}
\label{sec:linearPA}

In this section, calculations are performed and illustrated under the assumption that the reference signal for RF cancellation is taken after the PA, according to Case A. Thus, the nonlinear distortion produced by the PA is included in the RF reference signal and consequently attenuated by the assumed amount of RF cancellation.

\subsubsection{Fixed Amount of Digital Cancellation}

\begin{figure}[!t]
\centering
\includegraphics[width=\columnwidth]{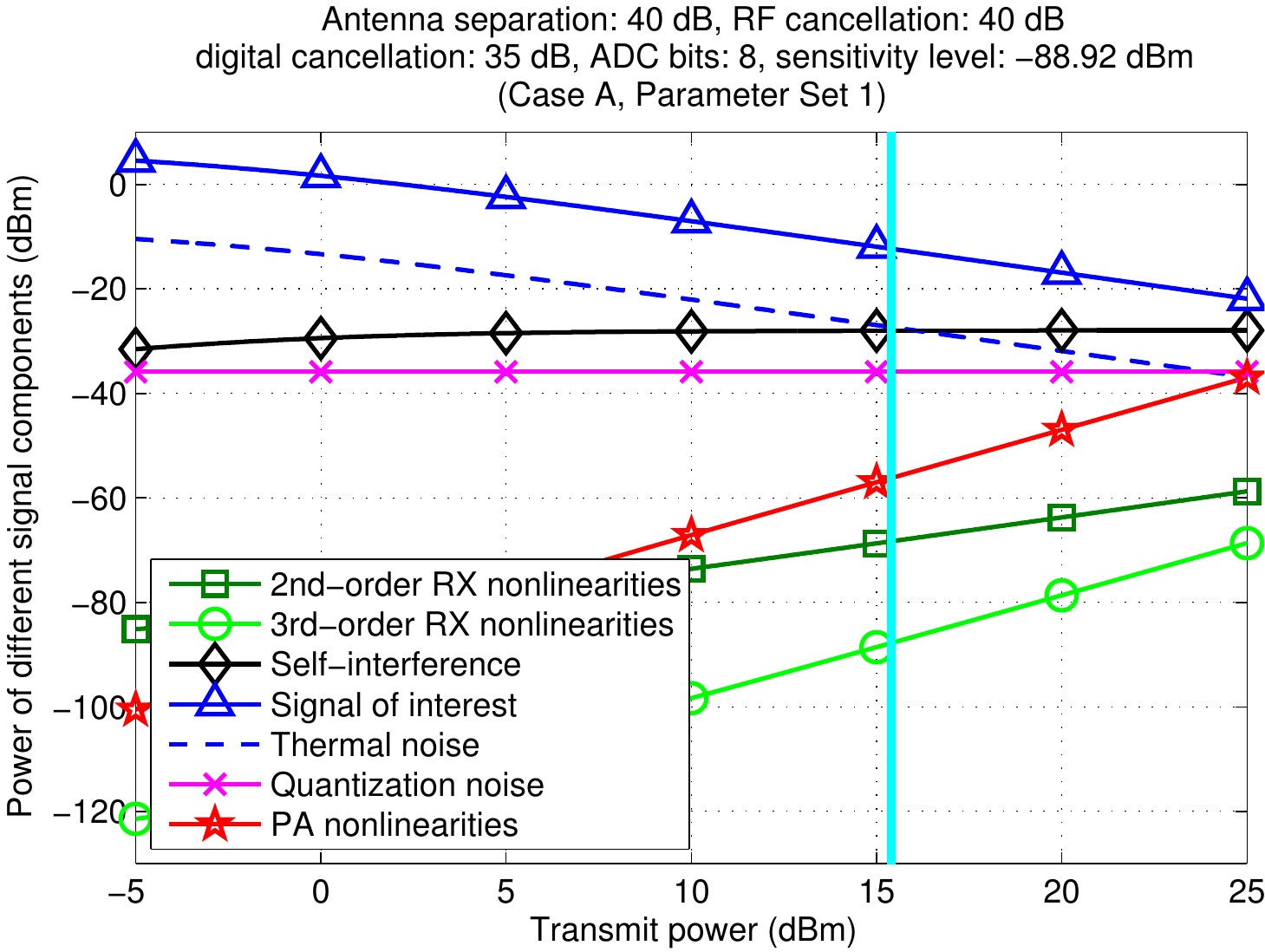}
\caption{The power levels of different signal components at the input of the detector with Parameter Set 1, assuming Case A.}
\label{fig:p_all_default}
\end{figure}

In the first part of the analysis, Parameter Set 1 is used and only the transmit power of the transceiver is varied, while all the other parameters remain constant and unaltered. The power levels of the different signal components can be seen in Fig.~\ref{fig:p_all_default} in terms of transmit power. The power levels have been calculated using (\ref{eq:p_quant})--(\ref{eq:p3}) with the selected parameters. It is imminently obvious that with the chosen parameters, the actual SI is the most significant distortion component. Furthermore, it can be observed that the maximum transmit power is approximately 15 dBm, marked by a vertical line. After this point, the loss of SINR due to SI becomes greater than 3 dB because the SI becomes equally powerful as thermal noise. When interpreting the behavior of the curves in Fig.~\ref{fig:p_all_default}, one should also remember that the power of the signal entering the ADC is kept approximately constant by the AGC. Thus, in practise, the total gain of the RX chain reduces when transmit power increases.

\begin{figure}[!t]
\centering
\includegraphics[width=\columnwidth]{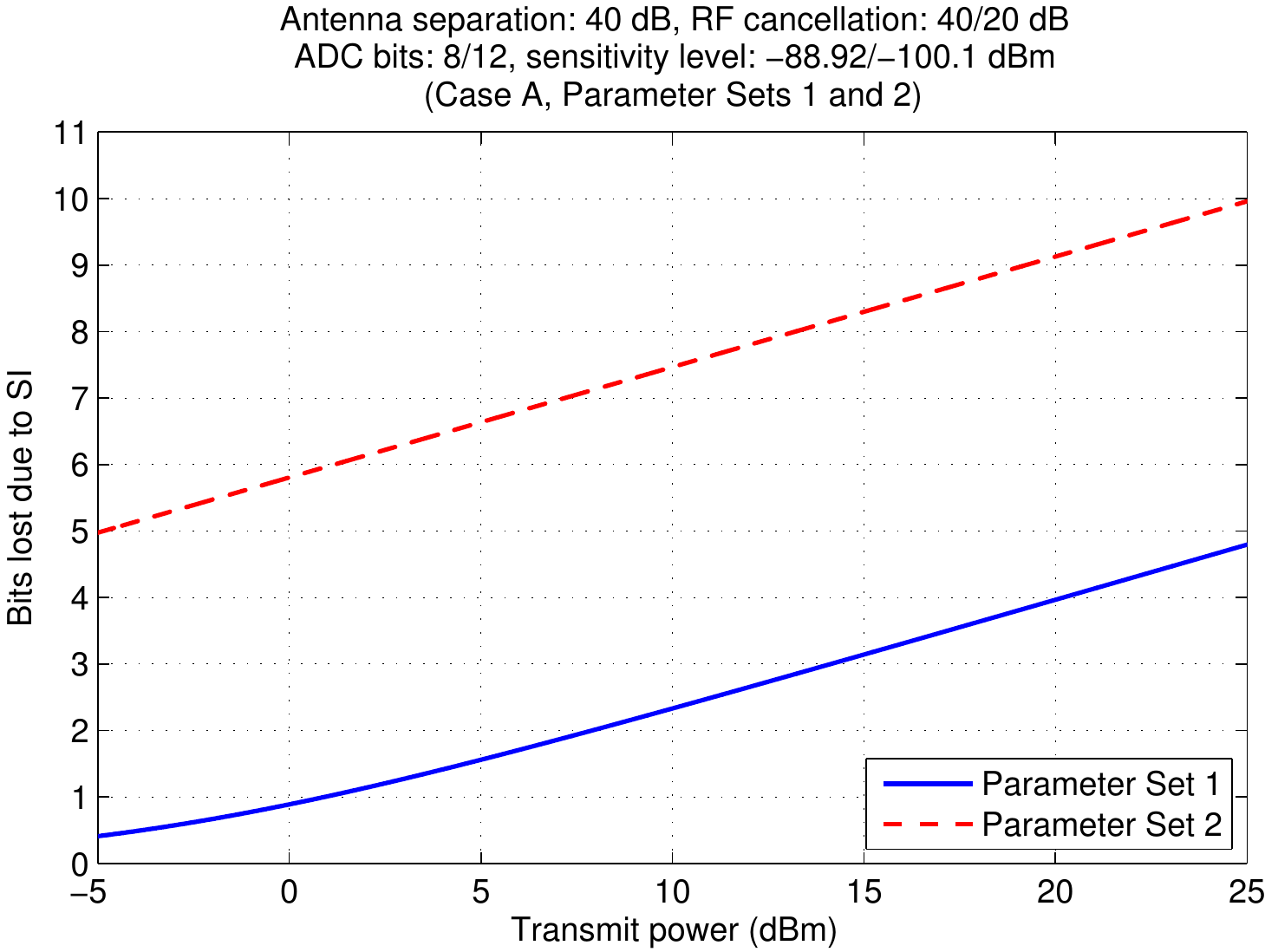}
\caption{The amount of lost bits due to SI with both parameter sets in Case A.}
\label{fig:bitloss_default}
\vspace{-3.5mm}
\end{figure}

The amount of lost bits, with respect to transmit power, can be seen in Fig.~\ref{fig:bitloss_default}. The curve is calculated with (\ref{eq:bitloss}) and it tells how much of the dynamic range of the ADC is effectively reserved by SI. It can be observed that when using Parameter Set 1, approximately 3 bits are lost due to SI with the maximum transmit power of 15 dBm. This emphasizes the fact that, in this scenario, the actual SI is the limiting factor for the transmit power. Actually, the power of quantization noise is almost 10 dB lower. However, from Fig.~\ref{fig:bitloss_default} it can also be observed that, with a transmit power of 20 dBm, the bit loss is already 4 bits. \emph{This indicates that, in order to enable the usage of higher transmit powers, a high-bit ADC is required}.

\subsubsection{Variable Amount of Digital Cancellation and Pushing the Performance Limits}

In order to further analyze the limits set by the analog-to-digital conversion and nonlinear distortion, it is next assumed that the amount of digital linear cancellation can be increased by an arbitrary amount, while the other parameters remain constant. With this assumption, it is possible to cancel the remaining linear SI perfectly in the digital domain. The reason for performing this type of an analysis is to determine the boundaries of DSP-based SI cancellation, as it would be beneficial to cancel as large amount of SI in the digital domain as possible. However, in many cases, increasing only digital cancellation is not sufficient to guarantee a high enough SINR because nonlinear distortion and quantization noise increase the effective noise floor above the allowed level.

To observe these factors in more detail, the amount of digital cancellation is next selected so that the loss of SINR caused by SI is fixed at 3 dB. This means that the combined power of the other distortion components is allowed to be equal to the power of the thermal noise included in the received signal. Thus, in this case, if the ratio between the signal of interest and dominating distortion becomes smaller than 15 dB, the above condition does not hold, and the loss of SINR becomes greater than 3 dB.

Below we provide closed-form solution for the  required amount of digital cancellation. The linear SINR requirement, which must be fulfilled after digital cancellation, is denoted by $\mathit{sinr}_{\text{RQ}} $. Then, the SINR requirement can only be fulfilled if
\begin{align} \label{eq:sinr_rq}
\mathit{sinr}_{\text{RQ}} < \frac{g_\text{rx}p_{\text{SOI,in}}}{g_\text{rx} Fp_{\text{N,in}}+p_{\text{2nd}}+p_{\text{3rd}}+\frac{g_\text{rx} p_{\text{3rd,PA,tx}}}{a_{\text{ant}} a_\text{RF}}+p_{\text{quant}}} \text{.}
\end{align}
In words, the SINR must be above the minimum requirement without taking the SI into account. If it is assumed that the above condition holds, the required SINR can be achieved with digital cancellation, and it can be written as
\begin{align} \label{eq:sinr_rq_2}
\mathit{sinr}_{\text{RQ}} = \frac{g_\text{rx}p_{\text{SOI,in}}}{\splitfrac{g_\text{rx}Fp_{\text{N,in}}+ \frac{g_\text{rx}}{a_{\text{ant}}}\left( \frac{p_{\text{tx}}}{a_{\text{RF}} a_\text{dig}} + \frac{p_{\text{3rd,PA,tx}}}{a_{\text{RF}}} \right) +p_{\text{2nd}}+\cdots}{+p_{\text{3rd}}+p_{\text{quant}}}} \text{.}
\end{align}
From here, the amount of required digital cancellation can be solved and written as
\begin{align} \label{eq:a_dig}
a_{\text{dig}} &= \frac{\frac{g_\text{rx} p_{\text{tx}}}{a_{\text{ant}}a_{\text{RF}}}}{\frac{g_\text{rx} p_{\text{SOI,in}}}{\mathit{sinr}_{\text{RQ}}}-(g F p_{\text{N,in}}+p_{\text{2nd}}+p_{\text{3rd}}+\frac{g_\text{rx} p_{\text{3rd,PA,tx}}}{a_{\text{ant}} a_\text{RF}}+p_{\text{quant}})}\nonumber\\
&= \frac{1}{1+\frac{a_{\text{ant}}a_{\text{RF}}p_{\text{SOI,in}}}{p_{\text{tx}}}\left(\frac{1}{\mathit{sinr}_{\text{RQ}}}-\frac{1}{sinr_{\text{DC}}}\right)}  \text{,}
\end{align}
where $sinr_{\text{DC}}$ is the linear SINR before digital cancellation. The first form of the equation above shows that the amount of required digital cancellation depends directly on the transmit power. It can also be observed that increasing antenna separation or RF cancellation decreases the requirements for digital cancellation.

\begin{figure}[!t]
\centering
\includegraphics[width=\columnwidth]{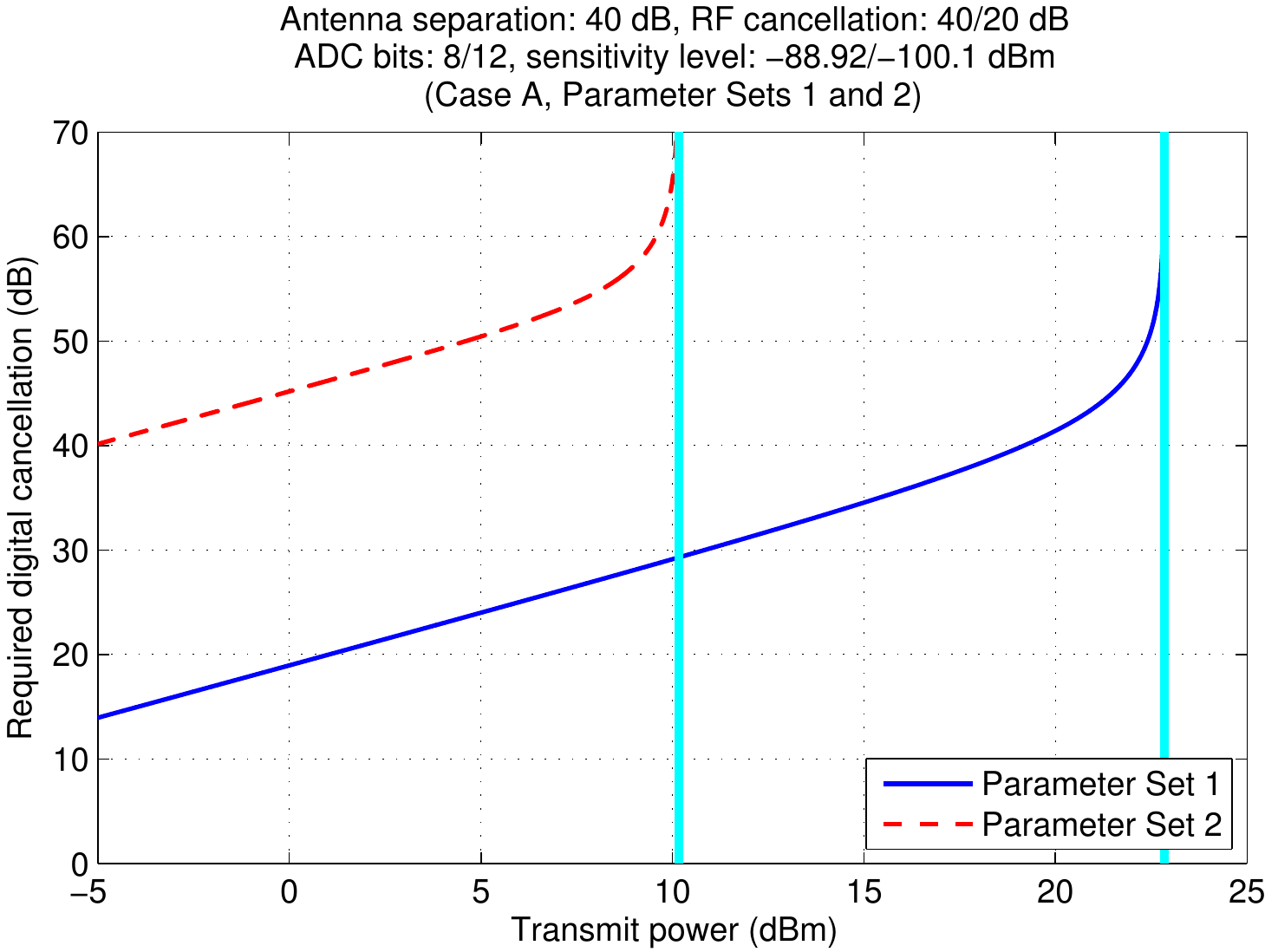}
\caption{The required amount of digital cancellation to sustain a 3 dB SINR loss with both parameter sets in Case A.}
\label{fig:digCanc}
\vspace{-3.5mm}
\end{figure}

The required amount of digital cancellation to sustain a maximum of 3 dB SINR loss, calculated from (\ref{eq:a_dig}), is illustrated in Fig.~\ref{fig:digCanc} in terms of the transmit power. The other parameters, apart from digital cancellation, are kept constant. It can be observed from the figure that the maximum transmit power is approximately 23 dBm for Parameter Set 1. After this, the amount of required digital cancellation increases to infinity, indicating perfect linear SI cancellation. However, as discussed earlier, after this point even perfect linear digital cancellation is not sufficient to maintain the required SINR, because quantization noise and nonlinearities become the limiting factor.

\begin{figure}[!t]
\centering
\includegraphics[width=\columnwidth]{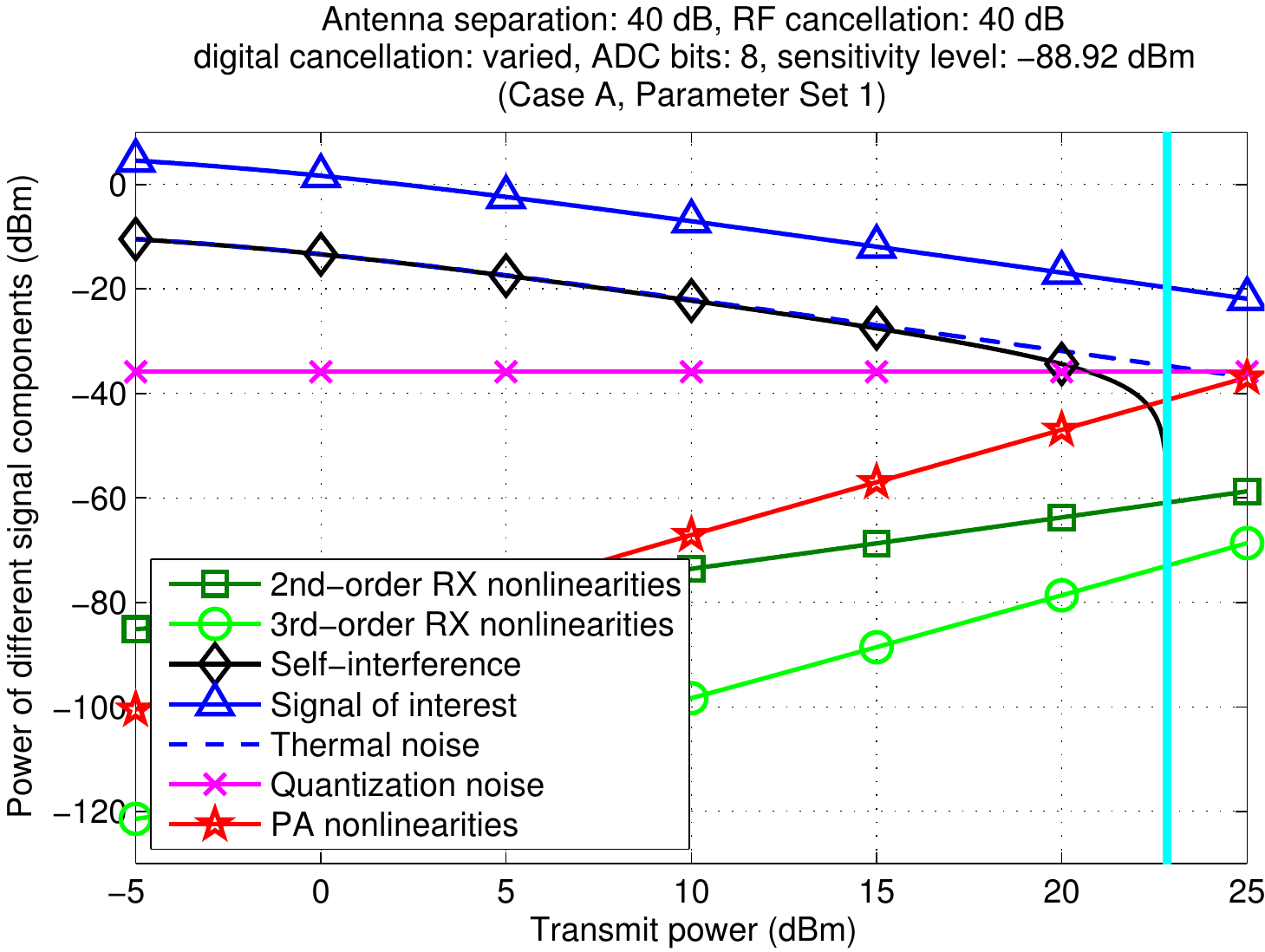}
\caption{The power levels of different signal components at the input of the detector when the amount of digital cancellation is increased, assuming Parameter Set 1 and Case A.}
\label{fig:P_all_default_incDC}
\vspace{-3mm}
\end{figure}

The power levels of the different signal components in this scenario are presented in Fig.~\ref{fig:P_all_default_incDC}. It can be observed that now quantization noise is the limiting factor for the SINR. The reason for this is that, with higher transmit powers and variable digital cancellation, the majority of SI is now cancelled in the digital domain and thus SI occupies the majority of the dynamic range of the ADC. This, on the other hand, deteriorates the resolution of the desired signal.

In order to further analyze the maximum allowed transmit power of the considered full-duplex transceiver, it is next determined how different parameters influence it. If we denote the signal-to-(thermal)noise-ratio at the detector by $\mathit{snr}_{\text{d}}$, the following equation holds when the loss of SINR is 3 dB:
\begin{align}
\mathit{snr}_{\text{d}} = \frac{g_\text{rx} p_{\text{SOI,in}}}{\frac{g_\text{rx}}{a_{\text{ant}}}\left( \frac{p_{\text{tx,max}}}{a_{\text{RF}} a_\text{dig}} + \frac{p_{\text{3rd,PA,tx}}}{a_{\text{RF}}} \right)+p_{\text{2nd}}+p_{\text{3rd}}+p_{\text{quant}}} \text{.} \label{eq:snr}
\end{align}
This means that the total power of the other types of distortion is equal to the thermal noise power, resulting in 3 dB SINR loss. When considering the maximum transmit power, it is again assumed that digital SI cancellation is perfect. Furthermore, as the transmit power is high, and also the nonlinear distortion produced by the PA is attenuated by RF cancellation, the power of SI can be used to approximate the power of the total signal at the input of the RX chain. This, on the other hand, allows us to use (\ref{eq:gain_appr}) to approximate the total receiver gain. Thus, when substituting $g_\text{rx}$ with (\ref{eq:gain_appr}), letting $ a_{\text{dig}} \to \infty $, and expressing quantization noise as $\frac{p_{\text{target}}}{\mathit{snr}_{\text{ADC}}} $, (\ref{eq:snr}) becomes
\begin{align}
\mathit{snr}_{\text{d}} &= \frac{\frac{a_{\text{ant}} a_{\text{RF}} p_{\text{target}}}{p_{\text{tx,max}}} p_{\text{SOI,in}}}{ \frac{a_{\text{ant}} a_{\text{RF}} p_{\text{target}}}{p_{\text{tx,max}}} \frac{p_{\text{3rd,PA,tx}}}{a_\text{ant} a_\text{RF}} + p_{\text{2nd}}+p_{\text{3rd}}+\frac{p_{\text{target}}}{\mathit{snr}_{\text{ADC}}}}\nonumber\\ 
&=\frac{a_{\text{ant}} {a_\text{RF}} p_{\text{SOI,in}}}{p_{\text{tx,max}} \left(\frac{p_{\text{2nd}}+p_{\text{3rd}}}{p_{\text{target}}}+\frac{1}{\mathit{snr}_{\text{ADC}}}\right) + p_{\text{3rd,PA,tx}}}  \text{.} \label{eq:snr_general}
\end{align}
By solving (\ref{eq:snr_general}) in terms of $p_{\text{tx,max}}$, the maximum transmit power can be obtained. However, as the power of nonlinear distortion is dependent on the transmit power, it is not convenient to derive an analytical equation for the maximum transmit power as it would require solving the roots of a 3rd-order polynomial. On the other hand, if we consider the scenario of Fig.~\ref{fig:P_all_default_incDC}, it can be seen that the quantization noise is actually the dominant distortion component. Thus, in this case, $p_{\text{2nd}}+p_{\text{3rd}} \approx 0 $ and $p_\text{3rd,PA,tx} \approx 0$, and the maximum transmit power becomes
\begin{align}
 p_{\text{tx,max}} &= \frac{a_{\text{ant}} {a_\text{RF}} p_{\text{SOI,in}} \mathit{snr}_{\text{ADC}}}{\mathit{snr}_{\text{d}}}\nonumber\\
\text{i.e. } P_{\text{tx,max}} &= A_{\text{ant}} + A_{\text{RF}} + P_{\text{SOI,in}} + \mathit{SNR}_{\text{ADC}} - \mathit{SNR}_{\text{d}}  \text{.} \label{eq:ptxmax}
\end{align}
By substituting $\mathit{SNR}_{\text{ADC}}$ with (\ref{eq:adc_snr}), we can approximate the maximum transmit power of the considered full-duplex transceiver as
\begin{align}
P_{\text{tx,max}} &= A_{\text{ant}} + A_{\text{RF}} + P_{\text{SOI,in}} - \mathit{SNR}_{\text{d}} + 6.02 b \nonumber\\
&- \mathit{PAPR} + 4.76 \text{.} \label{eq:maxtx}
\end{align}
This applies accurately when the quantization noise is the limiting factor.

An alternative possible scenario is the situation where the amount of bits is sufficiently high such that the quantization noise is not the main performance bottleneck. In this case, the power of nonlinear distortion is the limiting factor for the maximum transmit power (still assuming $ a_{\text{dig}} \to \infty $). In other words, if we let $ \mathit{snr}_{\text{ADC}} \to \infty $, \eqref{eq:snr_general} becomes
\begin{align}
\mathit{snr}_{\text{d}} = \frac{a_{\text{ant}} {a_\text{RF}} p_{\text{SOI,in}}}{p_{\text{tx,max}}\left(\frac{ p_{\text{2nd}}+p_{\text{3rd}}}{p_{\text{target}}}\right)+p_\text{3rd,PA,tx}}  \text{.} \label{eq:snr_infbits}
\end{align}
However, similar to solving (\ref{eq:snr_general}), it is again very inconvenient to derive a compact form for the maximum transmit power in this scenario, since it would again require solving the roots of a third order polynomial. Nevertheless, the value for the maximum transmit power can in this case be easily calculated numerically, which yields $p_{\text{tx,max}} \approx 25.02 \text{ dBm}$ and $p_{\text{tx,max}} \approx 10.29 \text{ dBm}$ with Parameter Sets 1 and 2, respectively.

\begin{figure}[!t]
\centering
\includegraphics[width=\columnwidth]{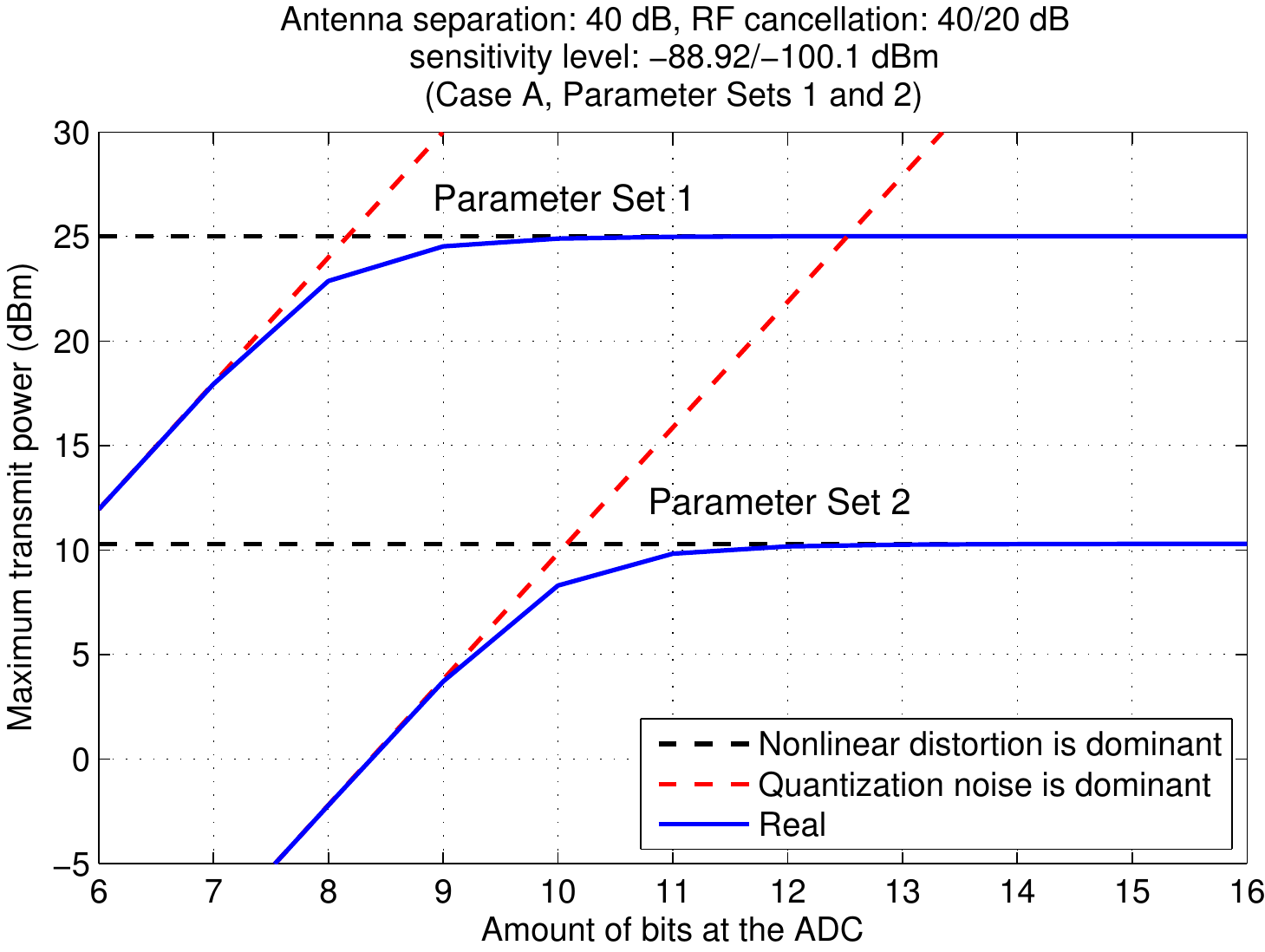}
\caption{The maximum transmit power with respect to the number of bits at the ADC, again with both parameter sets. The blue curve shows the real value of the maximum transmit power, and the red and black curves show the values when quantization noise or nonlinear distortion is the dominant distortion, respectively.}
\label{fig:maxTxPower}
\vspace{-3mm}
\end{figure}

If operating under such conditions that neither intermodulation nor quantization noise is clearly dominating, previous results in \eqref{eq:maxtx} and \eqref{eq:snr_infbits} may be overestimating the performance. For this reason, Fig.~\ref{fig:maxTxPower} shows the actual maximum transmit power with respect to the number of bits at the ADC without any such assumptions, calculated numerically from (\ref{eq:snr_general}). Also the maximum transmit powers for the two special scenarios are shown. With a low number of bits, the quantization noise is indeed the limiting factor for the transmit power, and the curve corresponding to (\ref{eq:maxtx}) is very close to the real value. On the other hand, with a high number of bits, the line corresponding to \eqref{eq:snr_infbits} is closer to the real value, as the power of quantization noise becomes negligibly low. This demonstrates very good accuracy and applicability of the derived analytical results.

Perhaps the most interesting observation from Fig.~\ref{fig:maxTxPower} is that, with Parameter Set 1, it is sufficient to have a 10 bit ADC in order to decrease the power of quantization noise negligibly low. This is shown by the fact that after that point, the maximum transmit power saturates to the value calculated with \eqref{eq:snr_infbits}. The saturated value of the maximum transmit power can only be increased by implementing more linear transceiver components or by increasing the amount of SI attenuation in the analog domain, thereby decreasing the power of nonlinear distortion and thus lowering the overall noise floor.

Overall, with the chosen parameters for the receiver, the bottleneck during the full-duplex operation in Case A is the quantization noise, in addition to the actual SI. This is an observation worth noting, as performing as much SI cancellation in the digital domain as possible is very desirable, since it allows the construction of cheaper and more compact full-duplex transceivers with affordable and highly-integrated RF components. In addition, it is also observed that, with higher transmit powers, the nonlinear distortion produced by the PA of the transmitter is a considerable factor. If a cheaper and less linear PA is used, this nonlinear distortion starts to limit even more heavily the achievable performance of a full-duplex transceiver.

\subsubsection{Calculations with Parameter Set 2}

In order to analyze how using cheaper, and hence lower-quality, components affects the RX chain, some calculations are done also with Parameter Set 2. The values of the parameters are again listed in Tables~\ref{table:system_parameters} and \ref{table:rx_parameters}. The sensitivity of the receiver is improved by decreasing the bandwidth and SNR requirement, and the power of the received signal is also decreased accordingly. In addition, the amount of RF cancellation is now assumed to be only 20 dB. This has a serious effect on the bit loss and the requirements for the digital cancellation.

The only component, whose specifications are improved, is the ADC, as it is now chosen to have 12 bits. The reason for this is to preserve a sufficient resolution for the signal of interest in the digital domain, as the amount of lost bits is relatively high with these weaker parameters. The calculations are again carried out assuming that the amount of digital cancellation can be increased arbitrarily high.

\begin{figure}[!t]
\centering
\includegraphics[width=\columnwidth]{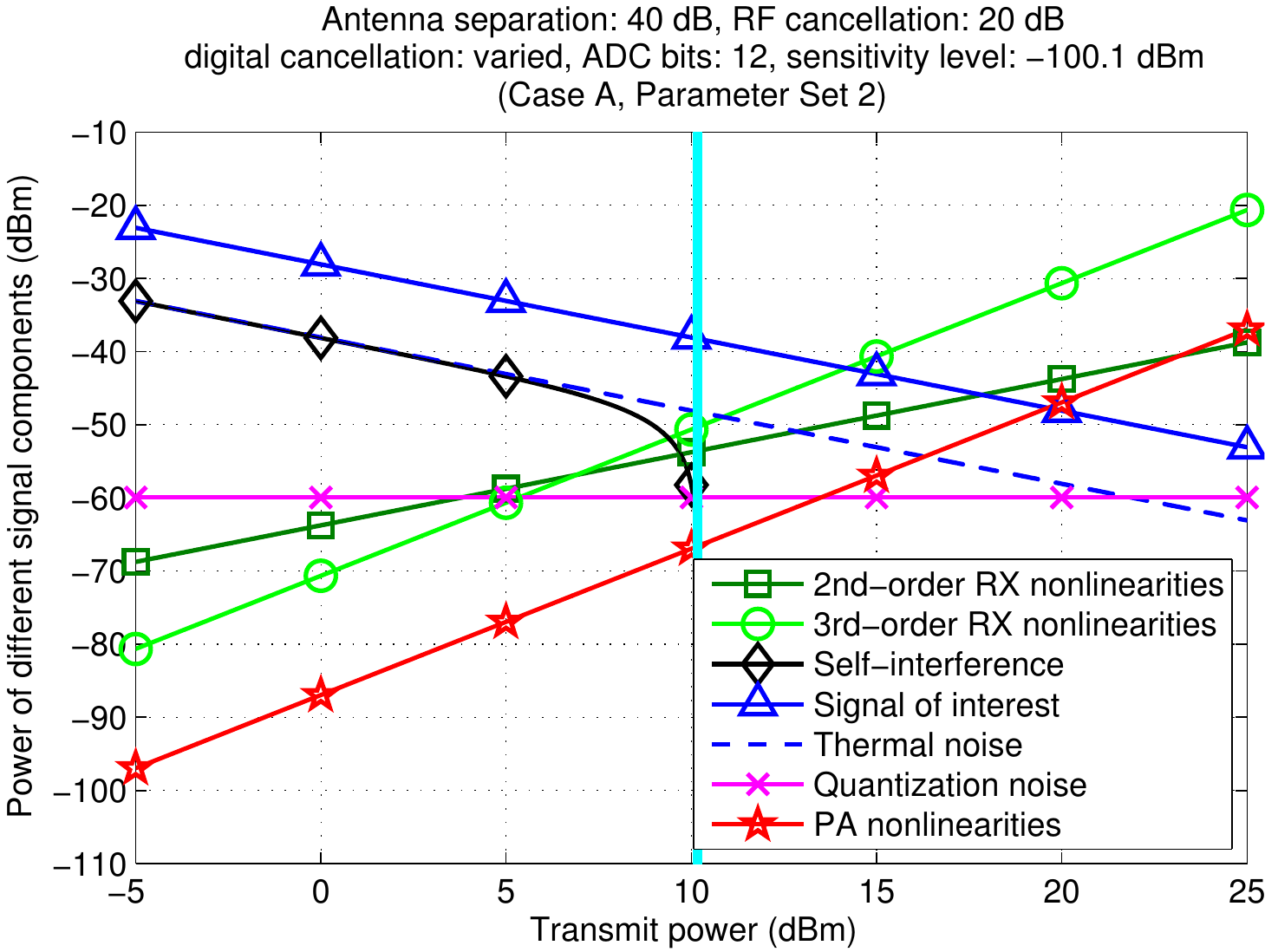}
\caption{The power levels of different signal components at the input of the detector with Parameter Set 2, assuming Case A.}
\label{fig:p_all_worsesens_incdc}
\end{figure}

The required amount of digital cancellation, when using Parameter Set 2, is depicted in Fig.~\ref{fig:digCanc}, and Fig.~\ref{fig:p_all_worsesens_incdc} shows the power levels of the different signal components in this scenario, again calculated with (\ref{eq:p_quant})--(\ref{eq:p3_pa}). It can be seen that now nonlinear distortion, produced by the receiver components, is the limiting factor for the transmit power, instead of quantization noise. The maximum transmit power is only approximately 10 dBm. After this point, mitigating only the linear SI is not sufficient to sustain the required SINR, as nonlinear distortion decreases the SINR below the required level.

With this parameter set, it can be seen that the amount of lost bits is very high (cf. Fig.~\ref{fig:bitloss_default}). This is due to the decreased RF cancellation ability, which means that the SI power is higher at the ADC interface. Thus, with lower SI cancellation performance at the analog/RF domain, the requirements for the ADC are heavily increased.

It can also be concluded that, with cheaper and less linear components, mitigating the RX chain nonlinear distortion with additional nonlinear DSP can provide performance gain. This is shown by Fig.~\ref{fig:maxTxPower}, where it can be observed that with Parameter Set~2, the maximum transmit power is decreased to 10~dBm, as opposed to the maximum transmit power of 25~dBm achieved with Parameter Set~1. This difference is caused by the lower linearity and decreased RF cancellation ability of the receiver utilizing Parameter Set~2. Thus, with decreased transceiver linearity and RF cancellation ability, also the nonlinear distortion produced by the RX chain must be considered, as Figs.~\ref{fig:maxTxPower} and~\ref{fig:p_all_worsesens_incdc} demonstrate.

\subsection{Results with Case B}
\label{sec:nonlinearPA}

In the system calculations of this section, Case B is considered, and thus the reference signal for RF cancellation is taken from the input of the PA. This means that the nonlinear distortion produced by the PA is not attenuated by RF cancellation, as it is not included in the cancellation signal. This obviously increases the effect of these TX-induced nonlinearities.

\begin{figure}[!t]
\centering
\includegraphics[width=\columnwidth]{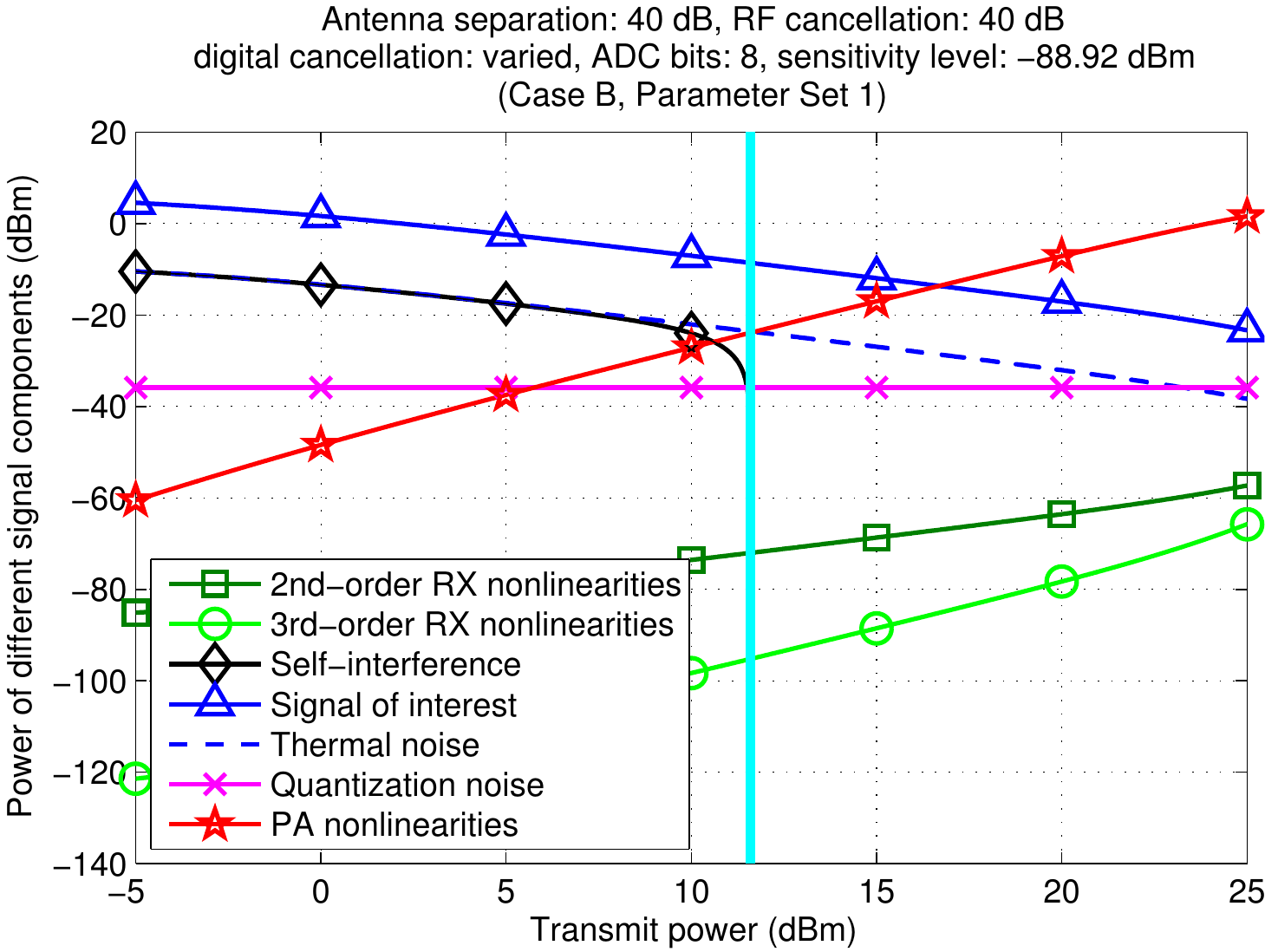}
\caption{The power levels of different signal components at the input of the detector with Parameter Set 1, assuming Case B.}
\label{fig:p_all_default_nlpa}
\end{figure}

The values for the parameters of the RX chain are chosen according to Parameter Set 1, and the amount of digital cancellation is again controlled to maintain a 3 dB loss of SINR. The transmit power is varied from $-$5~dBm to 25~dBm. Figure~\ref{fig:p_all_default_nlpa} illustrates the power levels of different signal components in this scenario. It can be observed that the nonlinear distortion produced by the PA is the most significant distortion component already with transmit powers higher than 11~dBm. Furthermore, with transmit powers higher than 12~dBm, it will decrease the SINR below the required level, thus preventing the usage of higher transmit powers.

When comparing Fig.~\ref{fig:p_all_default_nlpa} to Fig.~\ref{fig:P_all_default_incDC}, it can be observed that the difference is significant. This is caused by the fact that in Case B, the nonlinear distortion produced by the PA is not attenuated by RF cancellation, unlike in Case A. Hence, it is clear that an ability to mitigate nonlinear distortion would provide a significant performance gain for a full-duplex transceiver, which is implemented according to Case B. Furthermore, with the chosen parameters, it would be sufficient to mitigate the nonlinearities in the digital domain, as the quantization noise floor is fairly low relative to the other signal components.

\begin{figure}[!t]
\centering
\includegraphics[width=\columnwidth]{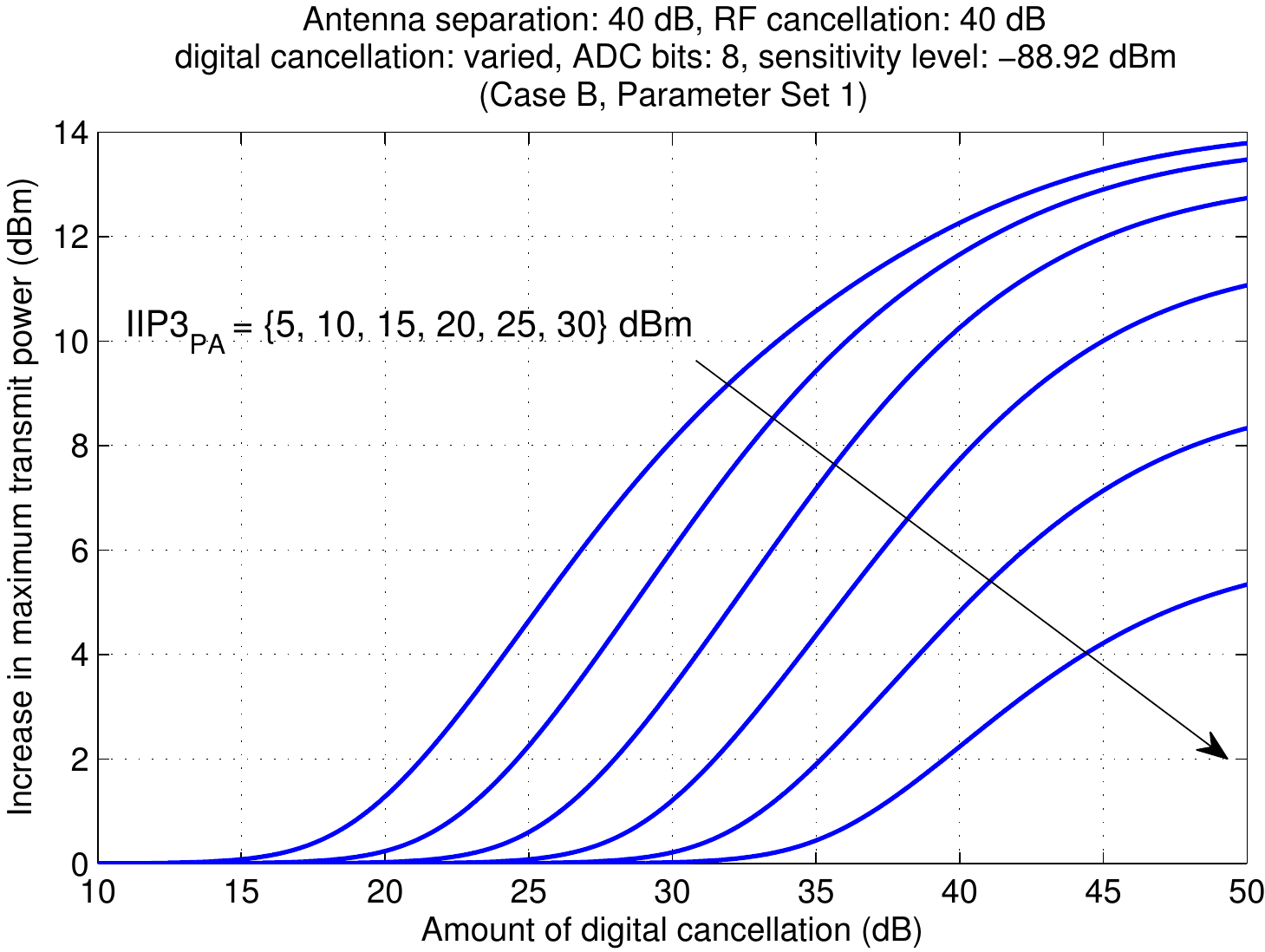}
\caption{The increase in maximum transmit power when also the nonlinear distortion of the SI channel can be mitigated with digital cancellation, compared to only linear cancellation. Horizontal axis depicts the total amount of achieved digital cancellation. The curves correspond to different IIP3 figures of the PA.}
\label{fig:maxtxpower_nlpa}
\vspace{-3mm}
\end{figure}

In order to demonstrate the potential of nonlinear cancellation, the maximum transmit powers of two different scenarios are compared. In the first case, it is assumed that digital cancellation is linear, and can thus mitigate only the linear part of the SI signal. In the other case, it is assumed that digital cancellation is able to mitigate also the nonlinear part of the SI signal, in addition to the linear part. Figure~\ref{fig:maxtxpower_nlpa} shows the increase in the maximum transmit power, when comparing these two scenarios. The same curve has also been plotted with different IIP3 values for the PA.  The curves have been calculated based on \eqref{eq:p_quant}--\eqref{eq:p3_pa}, with the modification that in the other case, $P_\text{3rd,PA}$ is also attenuated by $A_\text{dig}$. It can be observed that being able to mitigate the nonlinear component of the SI signal in the digital domain provides a significant increase in the maximum transmit power when the total amount of digital cancellation is increased. This has also been observed with actual waveform simulations in \cite{Anttila13}.

It can also be observed that already with 25 dB of digital cancellation, the maximum transmit power is increased by as much as 5 dB, if also the nonlinear component of the SI signal is mitigated. Obviously, the achievable gain is smaller with a more linear PA, and this indicates that when the nonlinear component of the SI signal is weaker, linear digital cancellation might be sufficient. However, with a less linear PA, significant increase in the maximum transmit power can be achieved with nonlinear digital cancellation, almost regardless of the total amount of achieved cancellation.

Overall, Figs.~\ref{fig:p_all_default_nlpa} and \ref{fig:maxtxpower_nlpa} illustrate that nonlinear distortion produced by the TX PA is a significant issue in full-duplex transceivers, when the reference signal for RF cancellation is taken from the input of the PA. Furthermore, the ability to compensate it can significantly improve the performance of the transceiver. Thus, implementing nonlinear estimation and processing mechanisms for digital SI cancellation is an interesting topic for future research.

\section{Waveform simulations and comparisons}
\label{sec:waveform_simul}

In order to analyze and demonstrate the good accuracy of the used models and the system calculation results, a complete full-duplex waveform simulator is constructed. It emulates a similar direct conversion transceiver that is used in the analytical calculations, having the parameters corresponding to Parameter Set 1. Here, only Case A is considered for compactness.

The simulator is implemented with Matlab and Simulink, using SimRF component library. The simulated waveform is chosen to be an OFDM signal with parameters specified in Table~\ref{table:simul_param}. These parameters are in essence similar to WLAN specifications, and they are used for generating both the transmitted and received signals.

\begin{table}[!t]
\renewcommand{\arraystretch}{1.3}
\caption{Additional parameters for the waveform simulator.}
\label{table:simul_param}
\centering
\begin{tabular}{|c||c|}
\hline
\textbf{Parameter} & \textbf{Value}\\
\hline
Constellation & 16-QAM\\
\hline
Number of subcarriers & 64\\
\hline
Number of data subcarriers & 48\\
\hline
Guard interval & 16 samples\\
\hline
Sample length & 15.625 ns\\
\hline
Symbol length & 4 $\mu$s\\
\hline
Signal bandwidth & 12.5 MHz\\
\hline
Oversampling factor & 4\\
\hline
ADC bits & 12\\
\hline
\end{tabular}
\end{table}

The SI channel is assumed to be static and it consists of a main coupling component and three weak multipath components, which are delayed by one, three, and eight sample intervals in relation to the main component, respectively. This corresponds to a maximum delay of 125 ns. The delay of the main component is assumed to be negligibly small, as the distance between the antennas is typically very short. The average power difference between the main component and the multipath components is set to 45 dB, which is on the same range as values measured in \cite{Duarte12}.

In these simulations, a single-tap RF canceller is considered, as this corresponds to the most typical scenario currently used in the literature \cite{Choi10,Jain11}. Thus, in this scenario, RF cancellation attenuates only the main coupling component of the SI signal. In the simulator, also some delay, amplitude, and phase errors are included in the RF cancellation signal to achieve the desired amount of SI attenuation, and to model the cancellation process in a realistic manner.

\begin{figure}[!t]
\centering
\includegraphics[width=\columnwidth]{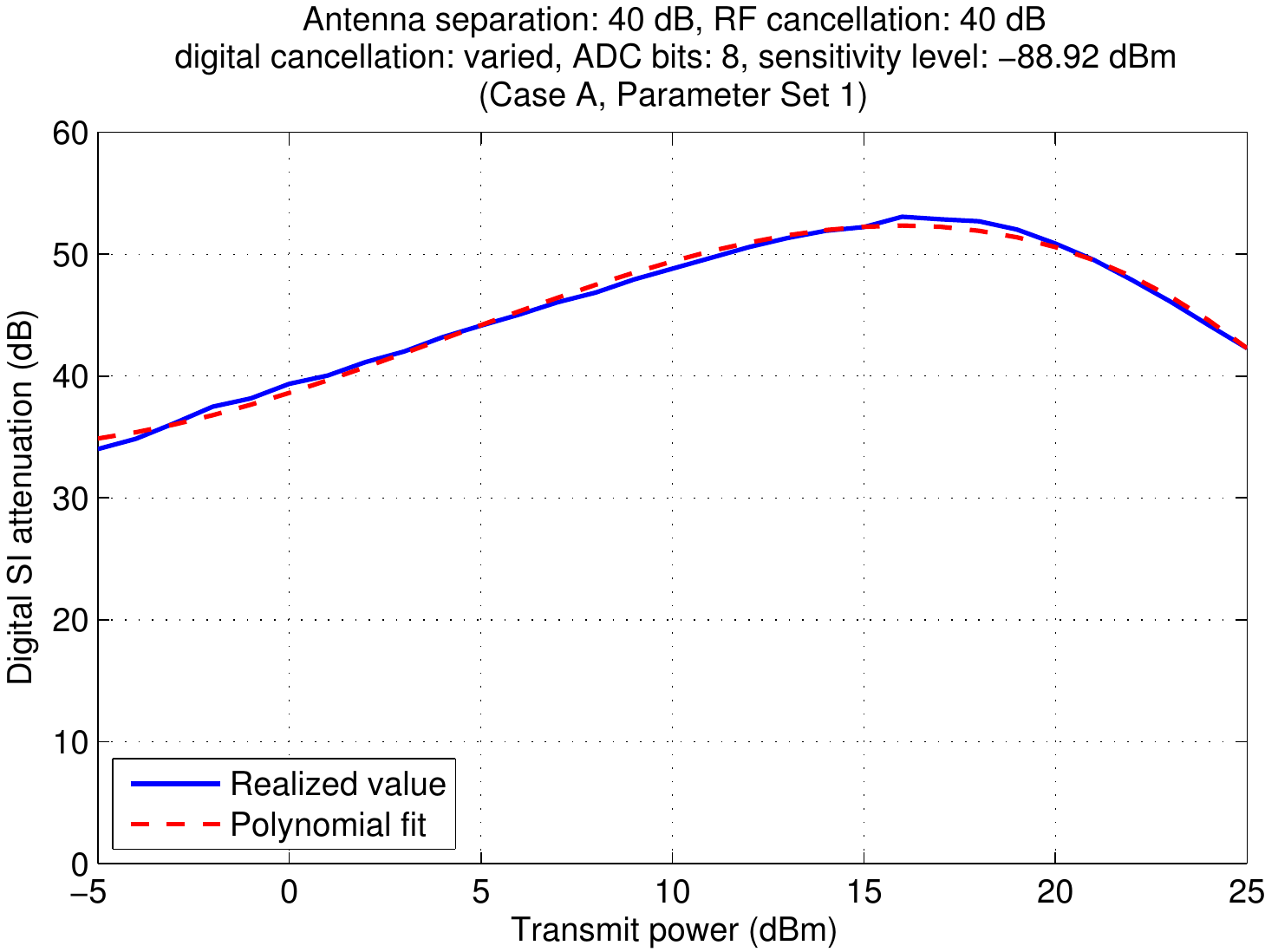}
\caption{The amount of achieved digital cancellation in the true waveform simulation, with respect to transmit power. TX-RX channel estimation in the digital cancellation is implemented with linear least-squares fitting between the ideal TX data and RX observation during a calibration period.}
\label{fig:dig_canc_simu}
\end{figure}

The attenuation of the weaker multipath components is then done by digital cancellation after the ADC. The implementation of digital cancellation utilizes classic least-squares based SI coupling channel estimation, which is implemented with linear least-squares fitting between the ideal TX data and RX observation during a calibration period. Thus, the amount of digital cancellation cannot be tuned arbitrarily since it depends directly on the accuracy of these TX-RX channel estimates. The amount of achieved digital cancellation is illustrated in Fig.~\ref{fig:dig_canc_simu}. The fluctuating curve is the realized value, and the smooth curve is a third order polynomial fitted to the realized values. The polynomial approximation is used when calculating the analytical SINR, in order to assess realistic average performance. As shown in Fig.~\ref{fig:dig_canc_simu}, larger amounts of cancellation are achieved with higher transmit powers, as the quality of the channel estimate is better with a stronger SI signal. This phenomenon has also been observed in practice \cite{Duarte12}. However, with transmit powers above 17 dBm, the power of the PA-induced nonlinear distortion starts to decrease the achievable digital cancellation.

\begin{figure}[!t]
\centering
\includegraphics[width=\columnwidth]{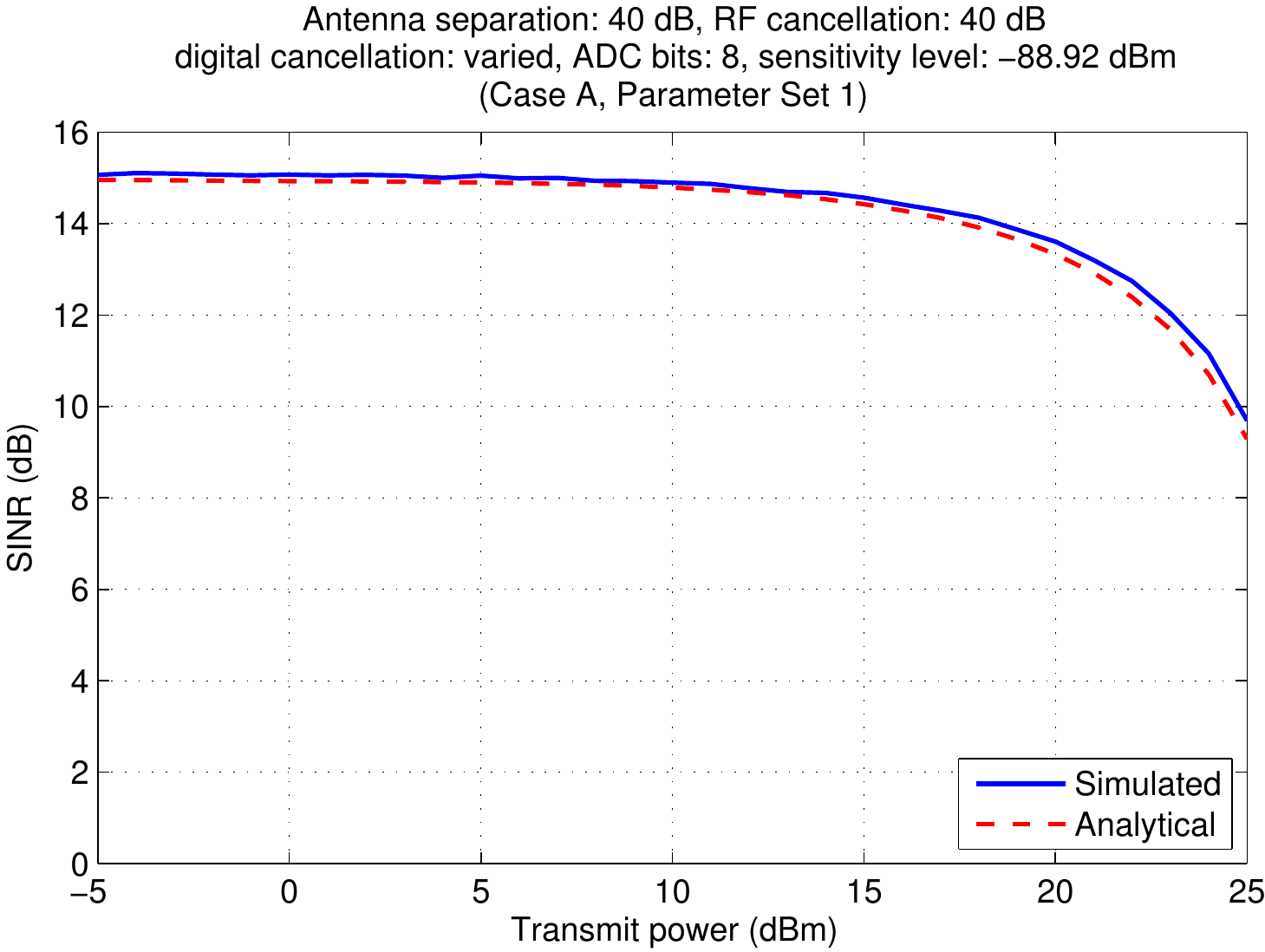}
\caption{SINR values obtained from the waveform simulations and from the analytical calculations.}
\label{fig:sinrloss_simu}
\end{figure}

The results of the analytical calculations are then compared to the simulation results in terms of the SINR at the input of the detector ($\mathit{SINR}_\text{d}$). Figure~\ref{fig:sinrloss_simu} shows the SINRs obtained with analytical calculations and with full waveform simulations, with respect to transmit power. In the waveform simulator, the SINR is calculated by first determining the effective powers for the ideal signal, and total noise-plus-interference signal. After this, the SINR is calculated as the ratio of these signal powers. The simulation is repeated 50 times for each transmit power, and the transmit power is varied with 1 dB intervals. The SINR corresponding to each transmit power is calculated as the average value of these independent realizations. The analytical SINR is calculated directly from the previously presented equations. From Fig.~\ref{fig:sinrloss_simu} it can be seen that the analytical and simulated SINR curves are practically identical, thus evidencing excellent accuracy and reliability of the reported analytical expressions. With closer inspection, it can be observed that the analytically calculated SINR is actually slightly pessimistic throughout the considered transmit power range, but the difference is only in the order of 0.1--0.3 dB. This is likely to be caused by the different approximations made when deriving the equations for the power levels of the different signal components. In any case, it can be concluded that the accuracy of the analysis is very high.

\section{Conclusion}
\label{sec:conclusions}

In this paper, the effects of self-interference on the receiver chain of a full-duplex transceiver were analyzed in detail, taking into accout realistic antenna isolation, RF cancellation and digital baseband cancellation. Specific emphasis was on modeling and analyzing the impacts of transmitter and receiver RF nonlinearities as well as analog-to-digital converter dynamic range requirements. The reliability of the analytical results was also verified and demonstrated by comparing them with the reference results acquired from complete full-duplex device waveform simulations. These comparisons showed excellent match, verifying the high accuracy and reliability of the results.

In terms of RF cancellation reference injection, the analysis covered two alternative scenarios where the reference is taken either from transmitter power amplifier output (Case A) or input (Case B). In Case A, it was observed that with high-quality RF components and RF cancellation, the maximum tolerable transmit power is mostly limited by the quantization noise of the receiver analog-to-digital converter, as well as by the achievable amount of linear digital cancellation. However, with low-cost receiver RF components and lower-quality RF cancellation, feasible for mobile devices, also the transceiver chain nonlinearities were found considerable and can actually become the limiting factor. In Case B, in turn, it was observed that the linearity of the power amplifier is the major bottleneck for the receiver performance with transmit powers above 10~dBm, even when clearly fulfilling any typical transmitter emission mask. This applies also to a closely-related architecture where separate low-power transmitter chain is used to generate the RF reference.

In order to be able to implement full-duplex transceivers with transmit powers in the order of 20--30 dBm, typical to WiFi and mobile cellular radio terminals, with low-cost RF electronics, the findings of this article strongly motivate for the development of nonlinear digital self-interference cancellation techniques. This applies to the 2nd- and 3rd-order inband nonlinear distortion of the receiver RF components, and in particular to the 3rd-order inband nonlinear distortion of the transmitter power amplifier. Developing such nonlinear cancellation techniques is the main topic of our future research.

\appendices

\section{Derivation of Bit Loss due to SI}
\label{app:bitloss}

A principal equation for the bit loss due to noise and interference is written in \eqref{eq:bitloss_basic}. However, as we are now interested in the amount of bits lost due to SI, the bit losses under HD and FD operation must be compared. By subtracting the amount of lost bits under HD operation from the amount of lost bits under FD operation, we obtain the desired value of bit loss due to SI, written as
\begin{align}
b_{\text{lost}} = \frac{P_{\text{target}}-P_{\text{SOI,FD}}}{6.02}-\frac{P_{\text{target}}-P_{\text{SOI,HD}}}{6.02} \text{,} \label{eq:bitloss_1}
\end{align}
where $P_{\text{target}}$ corresponds to the total power of the signal at the input of the ADC (which is always constant because of AGC), and $P_{\text{SOI,FD}}$ and $P_{\text{SOI,HD}}$ are the powers of the desired signal with and without SI, respectively. Because the total power of the signal at the input of the ADC is kept constant by the AGC, \eqref{eq:bitloss_1} can be further simplified to express the bit loss in terms of the gains as
\begin{align}\label{eq:bitloss_gain}
b_{\text{lost}} &= \frac{P_\text{SOI,HD}-P_\text{SOI,FD}}{6.02} \nonumber\\
&= \frac{P_\text{SOI,in}+G_\text{HD}-(P_\text{SOI,in}+G_\text{FD})}{6.02} = \frac{G_{\text{HD}}-G_{\text{FD}}}{6.02} \text{,}
\end{align}
where $G_{\text{FD}}$ is the total gain of the RX chain under FD operation, and $G_{\text{HD}}$ is the total gain under HD operation, correspondingly. This is a rather intuitive expression for the bit loss, as the power of SI is obviously included in $G_\text{FD}$ due the reduction of the gain by the AGC. Noting that $ G = P_\text{target}-P_\text{in} $ and $ 6.02 \approx 10\log_{10}(4)$, the bit loss can be written as
\begin{align}
b_{\text{lost}} &= \frac{(P_\text{target}-P_\text{in,HD})-(P_\text{target}-P_\text{in,FD})}{10\log_{10}(4)} \nonumber\\
&= \frac{P_\text{in,FD}-P_\text{in,HD}}{10\log_{10}(4)} = \frac{10\log_{10}\left(\frac{p_\text{in,FD}}{p_\text{in,HD}}\right)}{10\log_{10}(4)} = \log_{4}\left(\frac{p_\text{in,FD}}{p_\text{in,HD}}\right) \nonumber\\
&\approx \log_{4}\left(1 + \frac{p_\text{SI,in} + p_\text{3rd,PA,in}}{p_\text{SOI,in} + p_\text{N,in}}\right) \text{.} \label{eq:bitloss_deriv}
\end{align}
By denoting that $p_\text{SI,in} = \frac{p_\text{tx}}{a_\text{ant} a_\text{RF}} $ and $p_\text{3rd,PA,in} = \frac{p_\text{3rd,PA,tx}}{a_\text{ant}a_\text{NL}} = \frac{p_\text{tx}^3}{a_\text{ant}a_\text{NL} \mathit{iip3}_\text{PA}^2 g_\text{PA}^2}$, \eqref{eq:bitloss_deriv} can finally be written as
\begin{align}
b_{\text{lost}} = \log_{4} & \left[1 +\left( \frac{1}{p_\text{SOI,in} + p_\text{N,in}}\right) \right. \nonumber\\
&\left. {} \cdot \left(\frac{p_\text{tx}}{a_\text{ant} a_\text{RF}} + \frac{p_\text{tx}^3}{a_\text{ant} a_\text{NL} \mathit{iip3}_\text{PA}^2 g_\text{PA}^2}\right)\right] \text{.} \label{eq:bitloss_final}
\end{align}

\section{Derivations of Receiver Nonlinear Distortion Products}
\label{app:nonlinear}

The derivation of \eqref{eq:p2} and \eqref{eq:p3} is done based on the power of nonlinear distortion at the output of a single component. This, on the other hand, can be calculated with \eqref{eq:nonlinear}. In the considered full-duplex transceiver, only the mixer and the VGA produce SI-induced 2nd-order nonlinear distortion on to the signal band. However, all the components are assumed to produce 3rd-order nonlinear distortion.

The derivation is done with linear power units to present the calculations in a more compact form. The total power of the signal at the input of the RX chain is denoted as $ p_\text{in} $. It consists of the signal of interest, SI, and thermal noise. Furthermore, the increase in the thermal noise power occurring within the RX chain is omitted, as it has no significant effect on the power of the nonlinear distortion. Using \eqref{eq:nonlinear}, and expressing the output power in terms of gain and input power, the power of the 3rd-order nonlinear distortion at the output of the LNA can be written as
\begin{align}
P_\text{3rd,LNA} = G_\text{LNA}+P_\text{in}-2(\mathit{IIP3}_\text{LNA}-P_\text{in})
\end{align}
Using the corresponding linear units, this can be written as
\begin{align}
p_\text{3rd,LNA} = \frac{g_\text{LNA}p_\text{in}^3}{\mathit{iip3}_\text{LNA}^2}  \text{.} \label{eq:p3_lna}
\end{align}

Now, noting that with the chosen parameters the power of the nonlinear distortion is negligibly small in comparison to the total power of the signal, the input power of the mixer can be written as
\begin{align}
p_\text{in,mixer} = g_\text{LNA} p_\text{in} + p_\text{3rd,LNA} \approx g_\text{LNA} p_\text{in}  \text{.}
\end{align}
The power of the 2nd-order nonlinear distortion produced by the mixer can be then written as
\begin{align}
p_\text{2nd,mixer} &= \frac{g_\text{mixer}p_\text{in,mixer}^2}{\mathit{iip2}_\text{mixer}} = \frac{g_\text{mixer}}{\mathit{iip2}_\text{mixer}} \left(g_\text{LNA} p_\text{in} \right)^2\nonumber\\
& = \frac{g_\text{LNA}^2 g_\text{mixer} p_\text{in}^2}{\mathit{iip2}_\text{mixer}} \text{.} \label{eq:p2_mixer}
\end{align}
The power of the 3rd-order nonlinear distortion produced by the mixer can in turn be written as
\begin{align}
p_\text{3rd,mixer} &= \frac{g_\text{mixer}p_\text{in,mixer}^3}{\mathit{iip3}_\text{mixer}^2} = \frac{g_\text{mixer}}{\mathit{iip3}_\text{mixer}^2} \left(g_\text{LNA} p_\text{in}\right)^3\nonumber\\
&= \frac{g_\text{LNA}^3 g_\text{mixer} p_\text{in}^3}{\mathit{iip3}_\text{mixer}^2} \text{.} \label{eq:p3_mixer}
\end{align}

Again, noting that the power of the nonlinear distortion is negligibly small in comparison to the total power of the signal, the input power of the VGA can be written as
\begin{align}
p_\text{in,VGA} \approx g_\text{mixer} p_\text{in,mixer} = g_\text{LNA} g_\text{mixer} p_\text{in} \text{.} 
\end{align}
The power of the 2nd-order nonlinear distortion at the output of the VGA can thus be written as
\begin{align}
p_\text{2nd,VGA} &= \frac{g_\text{VGA}p_\text{in,VGA}^2}{\mathit{iip2}_\text{VGA}}= \frac{g_\text{VGA}}{\mathit{iip2}_\text{VGA}} \left(g_\text{LNA} g_\text{mixer} p_\text{in}\right)^2\nonumber\\ 
&= \frac{g_\text{LNA}^2 g_\text{mixer}^2 g_\text{VGA} p_\text{in}^2}{\mathit{iip2}_\text{VGA}} \text{.} \label{eq:p2_vga}
\end{align}
Similarly, the power of 3rd-order nonlinear distortion at the output of the VGA can be written as
\begin{align}
p_\text{3rd,VGA} &= \frac{g_\text{VGA}p_\text{in,VGA}^3}{\mathit{iip3}_\text{VGA}^2} = \frac{g_\text{VGA}}{\mathit{iip3}_\text{VGA}^2}\left(g_\text{LNA} g_\text{mixer} p_\text{in}\right)^3 \nonumber\\
&= \frac{g_\text{LNA}^3 g_\text{mixer}^3 g_\text{VGA} p_\text{in}^3} {\mathit{iip3}_\text{VGA}^2} \text{.} \label{eq:p3_vga}
\end{align}

Finally, the total power of the nonlinear distortion of a given order can be determined by summing the powers of the nonlinear distortion at the output of each individual component, including also the effect of the gains of the upcoming components. Thus, the total power of the 2nd-order nonlinear distortion can be written as follows, using \eqref{eq:p2_mixer} and \eqref{eq:p2_vga}:
\begin{align}
p_\text{2nd} &= g_\text{VGA} p_\text{2nd,mixer} + p_\text{2nd,VGA} \nonumber\\
&=g_\text{VGA} \frac{g_\text{LNA}^2 g_\text{mixer} p_\text{in}^2}{\mathit{iip2}_\text{mixer}} + \frac{g_\text{LNA}^2 g_\text{mixer}^2 g_\text{VGA} p_\text{in}^2}{\mathit{iip2}_\text{VGA}} \nonumber\\
&= g_{\text{LNA}}^2 g_{\text{mixer}} g_{\text{VGA}} p_{\text{in}}^2 \left(\frac{1}{\mathit{iip2}_{\text{mixer}}}+\frac{g_{\text{mixer}}}{\mathit{iip2}_{\text{VGA}}} \right) \text{.}
\end{align}
Similarly, the total power of the 3rd-order nonlinear distortion can be written as follows, using \eqref{eq:p3_lna}, \eqref{eq:p3_mixer}, and \eqref{eq:p3_vga}:
\begin{align}
p_\text{3rd} &= g_\text{mixer}g_\text{VGA}p_\text{3rd,LNA} + g_\text{VGA}p_\text{3rd,mixer} + p_\text{3rd,VGA}\nonumber\\
&= g_\text{mixer}g_\text{VGA}\frac{g_\text{LNA}p_\text{in}^3}{\mathit{iip3}_\text{LNA}^2} + g_\text{VGA}\frac{g_\text{LNA}^3 g_\text{mixer} p_\text{in}^3}{\mathit{iip3}_\text{mixer}^2}\nonumber\\ 
&+ \frac{g_\text{LNA}^3 g_\text{mixer}^3 g_\text{VGA} p_\text{in}^3} {\mathit{iip3}_\text{VGA}^2}\nonumber\\
&= g_{\text{LNA}} g_{\text{mixer}} g_{\text{VGA}} p_{\text{in}}^3 \left[\left(\frac{1}{\mathit{iip3}_{\text{LNA}}}\right)^2\right.\nonumber\\
&\left. {}+\left(\frac{g_{\text{LNA}}}{\mathit{iip3}_{\text{mixer}}}\right)^2+\left(\frac{g_{\text{LNA}} g_{\text{mixer}}}{\mathit{iip3}_{\text{VGA}}}\right)^2 \right] \text{.}
\end{align}

When comparing the values calculated with the obtained equations to the values calculated without approximations, it is observed that the error is in the order of $ 0.7 \text{ } \% $ with transmit powers above 5 dBm. Thus, the approximations which are made in the derivation process do not have any notable effect on the reliability of the equations with the chosen parameter range.



\ifCLASSOPTIONcaptionsoff
  \newpage
\fi



\bibliographystyle{./IEEEtran}
\bibliography{./IEEEabrv,./IEEEref}

\begin{IEEEbiography}[{\includegraphics[width=1in,height=1.25in,clip,keepaspectratio]{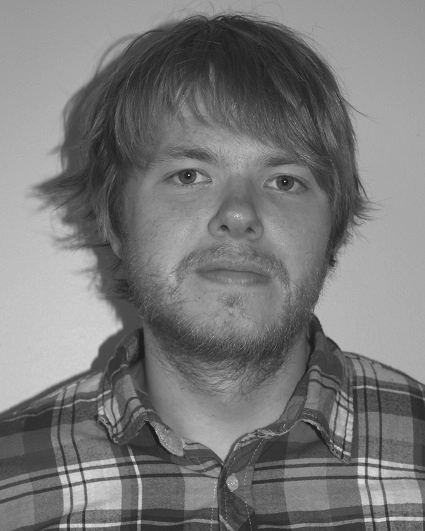}}]{Dani Korpi}
was born in Ilmajoki, Finland, on November 16, 1989. He received the B.Sc. degree (with honors) in communications engineering from Tampere University of Technology (TUT), Tampere, Finland, in 2012, and is currently pursuing the M.Sc. degree in communications engineering at TUT.

In 2011, he was a Research Assistant with the Department of Signal Processing at TUT. Since 2012, he has been a Research Assistant with the Department of Electronics and Communications Engineering, TUT. His main research interest is the study and development of single-channel full-duplex radios, with a focus on analysing the RF impairments.
\end{IEEEbiography}

\begin{IEEEbiography}[{\includegraphics[width=1in,height=1.25in,clip,keepaspectratio]{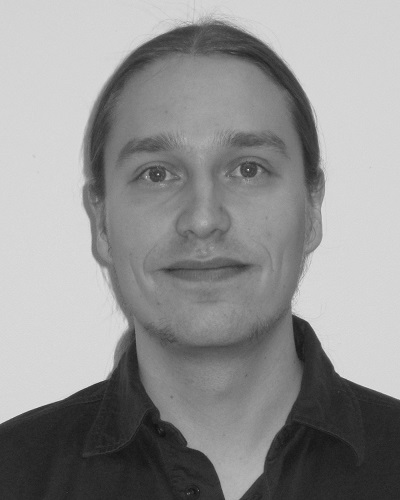}}]{Taneli Riihonen}
(S'06, M'14) received the M.Sc. degree in communications engineering (with distinction) from Helsinki University of Technology, Helsinki, Finland in February 2006.

During the summer of 2005, he was an intern at Nokia Research Center, Helsinki, Finland. Since fall 2005, he has been a researcher at the Department of Signal Processing and Acoustics, Aalto University School of Electrical Engineering, Helsinki, Finland, where he is completing his D.Sc. (Tech.) degree in the near future. He has also been a student at the Graduate School in Electronics, Telecommunications and Automation (GETA) in 2006-2010. His research activity is focused on physical-layer OFDM(A), multiantenna and relaying techniques.
\end{IEEEbiography}

\begin{IEEEbiography}[{\includegraphics[width=1in,height=1.25in,clip,keepaspectratio]{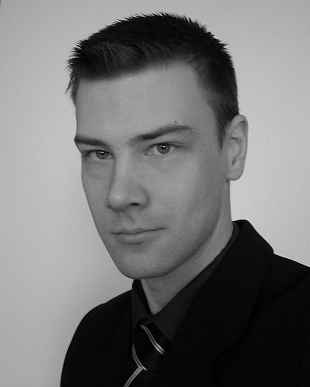}}]{Ville Syrj\"al\"a}
(S'09, M'12) was born in Lapua, Finland, in 1982. He received the M.Sc. (Tech.) degree in 2007 and D.Sc. (Tech.) degree in 2012 in communications engineering (CS/EE) from Tampere University of Technology (TUT), Finland.

He was working as a research fellow with the Department of Electronics and Communications Engineering at TUT, Finland, until 2013. Currently, he is working as a research fellow of the Japan Society for the Promotion of Science (JSPS) at Kyoto University, Japan. His general research interests are in full-duplex radio technology, communications signal processing, transceiver impairments, signal processing algorithms for flexible radios, transceiver architectures, direct sampling radios, and multicarrier modulation techniques.
\end{IEEEbiography}

\begin{IEEEbiography}[{\includegraphics[width=1in,height=1.25in,clip,keepaspectratio]{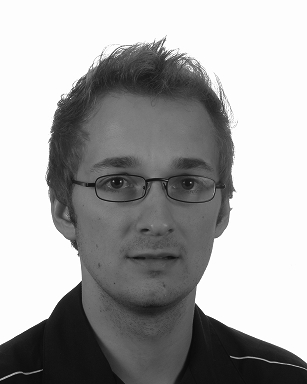}}]{Lauri Anttila}
(S'06, M'11) received his Ph.D. (with honours) from Tampere University of Technology (TUT), Tampere, Finland, in 2011.

Currently, he is a Research Fellow at the Department of Electronics and Communications Engineering at TUT. His current research interests include statistical and adaptive signal processing for communications, digital front-end signal processing in flexible radio transceivers, and full-duplex radio systems.
\end{IEEEbiography}

\begin{IEEEbiography}[{\includegraphics[width=1in,height=1.25in,clip,keepaspectratio]{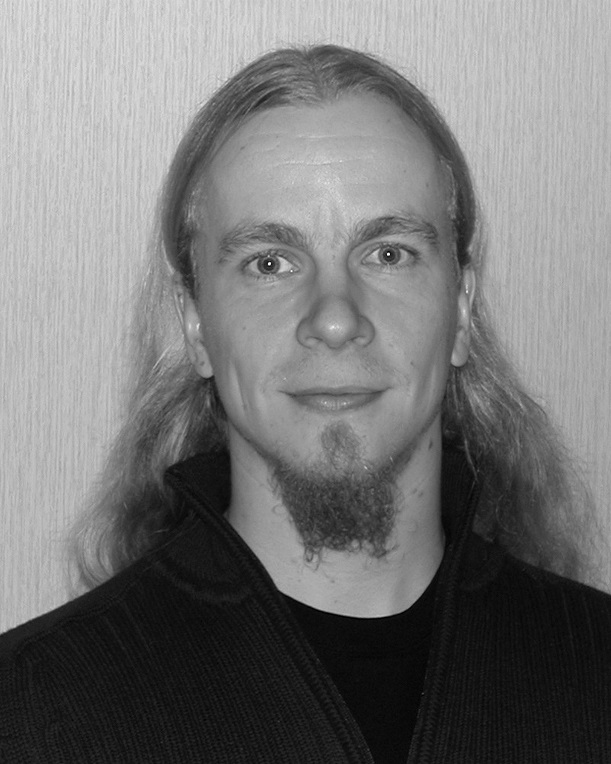}}]{Mikko Valkama}
(S'00, M'02) was born in Pirkkala, Finland, on November 27, 1975. He received the M.Sc. and Ph.D. degrees (both with honours) in electrical engineering (EE) from Tampere University of Technology (TUT), Finland, in 2000 and 2001, respectively. In 2002 he received the Best Ph.D. Thesis award by the Finnish Academy of Science and Letters for his dissertation entitled "Advanced I/Q signal processing for wideband receivers: Models and algorithms".

In 2003, he was working as a visiting researcher with the Communications Systems and Signal Processing Institute at SDSU, San Diego, CA. Currently, he is a Full Professor and Department Vice Head at the Department of Electronics and Communications Engineering at TUT, Finland. He has been involved in organizing conferences, like the IEEE SPAWC'07 (Publications Chair) held in Helsinki, Finland. His general research interests include communications signal processing, estimation and detection techniques, signal processing algorithms for software defined flexible radios, full-duplex radio technology, cognitive radio, digital transmission techniques such as different variants of multicarrier modulation methods and OFDM, radio localization methods, and radio resource management for ad-hoc and mobile networks.
\end{IEEEbiography}

\begin{IEEEbiography}[{\includegraphics[width=1in,height=1.25in,clip,keepaspectratio]{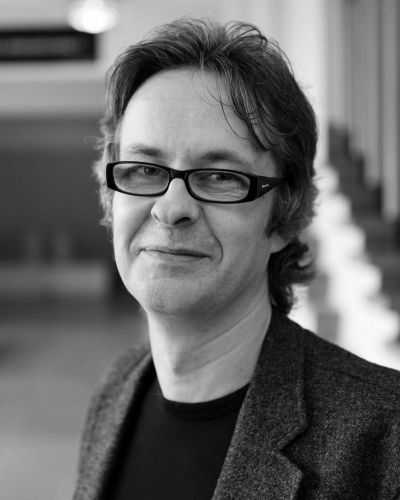}}]{Risto Wichman}
received his M.Sc. and D.Sc. (Tech) degrees in digital signal processing from Tampere University of Technology, Tampere, Finland, in 1990 and 1995, respectively.

From 1995 to 2001, he worked at Nokia Research Center as a senior research engineer. In 2002, he joined Department of Signal Processing and Acoustics, Faculty of Electrical Engineering, Aalto University, Finland, where he is a full professor since 2008. His research interests include signal processing techniques for wireless communication systems.
\end{IEEEbiography}

\end{document}